\documentclass[twocolumn,showpacs,preprintnumbers,amsmath,amssymb]{revtex4-2}
\usepackage{graphicx}
\usepackage{dcolumn}
\usepackage{color}
\usepackage{bm}
\usepackage{epsfig}
\usepackage{enumerate}
\usepackage{array}
\usepackage{multirow}
\usepackage{hyperref}
\usepackage{upgreek}
\begin{document}
\title{Quantum coherent control of nonlinear thermoelectric transport in a triple-dot Aharonov-Bohm heat engine}

\author{Jayasmita Behera,$^1$ Salil Bedkihal,$^2$ Bijay Kumar Agarwalla,$^3$ Malay Bandyopadhyay$^1$}
\affiliation{$^{1}$School of Basic Sciences, Indian Institute of Technology Bhubaneswar, Odisha, India 752050,\\
$^{2}$A.S.E Analytics India and 110 Rue St Francois N J1E 3H6, Sherbrooke, QC, Canada,\\
$^{3}$Department of Physics, Indian Institute of Science Education and Research, Pune 411008, India.}

\vskip-2.8cm
\date{\today}
\vskip-0.9cm

\begin{abstract}
%ABSTRACT
%\emph{Abstract version 1}
We investigate the role of quantum coherence and higher harmonics resulting from multiple-path interference in nonlinear thermoelectricity in a two-terminal triangular triple-dot Aharonov-Bohm (AB) interferometer. We quantify the trade-off between efficiency and power in the nonlinear regime of our simple setup comprising three non-interacting quantum dots (two connected to two biased metallic reservoirs) placed at the vertex of an equilateral triangle, and a magnetic flux $\Phi$ pierces it perpendicularly. For a spatially symmetric setup, we achieve optimal efficiency and power output when the inter-dot tunneling strength is comparable to the dot-lead coupling, AB phase $\phi=\pi/2$. Our analysis reveals that the presence of higher harmonics is necessary but not sufficient to achieve optimal power output. The maximal constructive interference represented by three close-packed resonance peaks of the unit transmission can enhance the power output ($P_{max}\sim 2.35\,\mathrm{fW}$) almost 3.5 times as compared to the case where only a single channel participates in the transport, and the corresponding efficiency is about $0.80\eta_{c}$ where $\eta_{c}$ is the Carnot efficiency. Geometric asymmetries and their effects on efficiency and power output are also investigated. An asymmetric setup characterized by the ratio of the coupling to the source and the drain terminals ($x$) can further enhance the maximum power output $P_{max}\sim 3.85\,\mathrm{fW}$ for $x=1.5$ with the same efficiency as that of the symmetric case. Our investigation reveals that the output power and efficiency are optimal in the wide-band limit. The power output is significantly reduced for the narrow-band case. On the other hand, disorder effects radically reduce the performance of the heat engine.
\end{abstract}
%below there are some pacs that we would typically use, pls search everytime to find most suitable
\pacs{85.25.Cp%Josephson devices
, 42.50.Dv %Quantum state engineering and measurements
%, 03.65.Ta%Measurement theory (quantum mechanics)
%, 85.25.Dq%SQUIDs
%03.67.Lx%Quantum computation
%07.57.Kp%microwave detectors~\cite{taylor_2008_NVDiamondMagn}
%85.25.Oj%superconducting photodetectors
}

\maketitle

%==========================
\section{Introduction}\label{intro}
Optimal conversion of heat to work is desirable for efficient energy harvesting technologies. The role of quantum coherence in energy harvesting in an efficient way is an ongoing research area. Recent advancements in nanotechnology have enabled us to study quantum-dot thermoelectric heat machines where quantum interference effects can play an important role. A major difference between these thermoelectric-based machines and other machines is that they do not have any macroscopic moving parts (i.e., no turbines, pistons, etc.). Rather, the working principles of such nanoscale thermoelectric devices are based on the steady-state currents of microscopic particles (electrons, phonons, etc.). The steady-state conversion of heat to work at the nanoscale has been reviewed in various works \cite{casati, engine1, Whitney2014, Whitney2015, Whitney2018, Taniguchi2020}.\\
\indent
In this perspective, nanoscale or low-dimensional materials are among the most encouraging candidates \cite{thermopower}. Typically, one considers a nanostructure made of a few quantum dots or a single-molecule junction between two thermal reservoirs maintained at different temperatures and electrochemical potentials. In such nanoscale devices, the striking role of symmetries, phase coherence, the effect of an external magnetic field, and quantum interference effects are investigated \cite{interf1,interf2,interf3,interf4,interf5,interf6,quantumphase1,quantumphase2,quantumphase3,quantumphase4,quantumphase5, Haack2019}. Numerous examples of heat engines using the quantum system as a working fluid have been proposed and even experimentally realized \cite{engine1,engine2,engine3}. One may mention the celebrated Aharonov-Bohm (AB) interferometer as a demonstrative example of a phase-tunable quantum device. The Aharonov-Bohm (AB) rings offer a tunable system to study the role of quantum effects in heat and charge transport \cite{ABring1, ABring2, ABring3,ABring4,ABring5}. For three-terminal quantum-dot heat engines using an AB interferometer, it has been shown that the introduction of magnetic flux can enhance heat-to-work conversion \cite{Wohlman, LuPRB100}. Quantum dot thermoelectric heat engines based on a two-terminal Aharonov-Bohm interferometer exhibiting sizable thermopower and the figure of merit exceeding one in the linear and nonlinear regimes have been studied \cite{Haack2019, Haack2021,Yamamoto1,Yamamoto2}.
The energy filtering mechanism introduced by quantum mechanics can lead to maximum efficiency where the system allows only particles from a specific energy window to pass \cite{Whitney2014, Whitney2015}.\\
%[The transmission function must be a sharply peaked boxcar function to obtain maximum efficiency. We find that the transmission function of approximate box car shape yet exhibits resonance peaks can deliver optimal power-efficiency configuration for moderate set-up asymmetries. The nonlinear thermoelectric transport in coupled nanostructures has been studied \cite{Hershfield2013}.] However, the interplay of internal coherent dynamics and coherent dot-lead coupling can give rise to a richer resonance structure. If the number of resonance peaks is equal to the number of transport channels such that each peak has unit transmission, then one can get enhanced power efficiency because of the maximal constructive interference. The role of such a higher-order interference pattern in thermoelectrics is relatively less explored.  \\
\indent
In this work, we study the triple quantum-dot Aharonov-Bohm (AB) interferometer as a thermoelectric heat engine. We consider a triangular geometry with magnetic flux threading the loop and two reservoirs connected at the terminals, termed as the source and drain. The source and drain are maintained at temperatures $T_S$ and $T_D$ and chemical potentials $\mu_S$ and $\mu_D$, respectively. To operate this setup as a thermoelectric heat engine, we maintain $T_S>T_D$ and $\mu_S<\mu_D$. In addition, the operation of the heat engine in the non-linear regime is controlled by considering $\Delta T=T_S-T_D\gg(T_S+T_D)/2$ and $\Delta\mu=\mu_D-\mu_S\gg(\mu_D+\mu_S)/2$. The steady-state behavior of the model has been studied at zero temperature limits \cite{BANDYOPADHYAY2021114786}. Most of the previous studies on AB interferometer heat engines are limited in the linear response regime \cite{engine1,engine2}. In the present study, we investigate the behavior of an AB interferometer operated in the fully non-linear regime as a quantum heat engine. We demonstrate the good tunability of the device (either by changing magnetic flux, or by controlling the ratio $\frac{t}{\gamma}$ with tunneling strength $t$ and coupling strength $\gamma$, or by changing the gate voltage), the sizable thermodynamic conversion efficiency, and large thermopower that this phase controllable thermoelectric quantum machine can attain. Although the controllability of the ratio $\frac{t}{\gamma}$ has been reported in previous investigations \cite{rogge,uditendu,chen,noiri,exptunits}, this corroborates the AB interferometer as an archetypal tool to develop efficient thermoelectric machines operating in the fully nonlinear quantum regime. Further, we compare the performance of our set-up between wide-band and narrow-band limits. We find that the narrow band case reduces the power output \cite{BANDYOPADHYAY2021114786,ABring4,ABring5}. Additionally, the analysis of the disorder effects due to fluctuation in the tunneling strengths as well as the on-site energies reveals a reduction in the performance of the heat engine by a significant amount.\\
\indent
Our investigation reveals that the quantum interference between the background continuum of states in the reservoirs and the discrete energy levels in the quantum dot leads to a formation of Fano resonance that is reflected as transmission asymmetries and sharp symmetric transmission resonances \cite{Hershfield2013,energy_filter, Clerk2001}. The role of Fano resonance in efficient heat-to-work conversion has been studied both theoretically and experimentally \cite{Fano1, Fano2, Fano3, Fano4, Fano5, Fano6}. Apart from the Fano resonance, the triangular geometric arrangement of quantum dots gives rise to electron trajectories winding around the loop multiple times that generate higher harmonics \cite{D'Anjou2013,harmonics1,harmonics2,harmonics3,harmonics4,harmonics5,harmonics6}. The effects of such higher-order interference patterns on the performance of thermoelectric heat engines are relatively less explored. The present work explores the role of higher-order harmonics (higher-order quantum interferences) in efficient heat-to-work conversion. At the outset we can summarize our main findings as follows: (i) The quantum interference effect can enhance the output power and efficiency of our AB thermoelectric heat engine. The optimal performance of our AB device can be achieved for the maximal constructive interference pattern which can be identified by the maximally compact transmission spectra (area under the transmission curve is required to be maximum such that all states contribute to the transport equally) of the system, (ii) This maximal constructive interference pattern can be achieved (as a result the power and efficiency of our phase tunable AB device can be optimized) by tuning external magnetic flux, and the ratio $\frac{t}{\gamma}$. We observe that the optimal power-efficiency can be achieved for magnetic flux $\phi=\frac{\pi}{2}$, and $\frac{t}{\gamma}=1$, (iii) We further find that the presence of anti-resonances, that gives rise to destructive interference channels reduces both the power output and efficiency. (iv) Tuning the asymmetric ratio of the system (the ratio of the coupling to the source and the drain terminals) can further enhance the performance of the engine. (v) The functioning of the AB heat engine is much superior in the wide-band limit compared to narrow-band limit. (vi) The disorder effects fanatically lower the power output of the engine.\\
\indent
With this background, the rest of the paper is organized as follows:
We first describe our model and the formalism in Sec.~\ref{sec2}. We analyze the density of states (DOS) and transmission function properties of our model set up in the fully non-linear regime in Sec.~\ref{occup}. Section \ref{power_harmonics} is devoted to the thermoelectric power and thermodynamic conversion efficiency of our setup and their tunability with magnetic flux, gate voltage, temperature bias, and dot-lead coupling strength. The connection between power-efficiency behavior and the harmonics of particle current and heat current for three different regimes demarcated by the parameter $t/\gamma$ ($t$ is the inter-dot tunneling strength and $\gamma$ is the hybridization strength) : (a) $t/\gamma < 1$ (b) $t/\gamma \sim 1$ and (c) $t/\gamma >1$. Our analysis shows that regime (b) with $\phi=\pi/2$ is the most optimal region for the operation of our thermoelectric engine. Section \ref{wbl} deals with the discussion of the thermoelectric response of our model by going beyond the wide-band limit approximation and the effects of disorders are discussed in Sec.~\ref{disorder}. In Sec.~\ref{geometry} we discuss the role of the geometry of the triple-dot setup for enhanced thermoelectric response. We conclude our paper in Sec.~\ref{conclusion}. We quantify the performance of the thermoelectric heat engine by investigating the Seebeck coefficient in the appendix \ref{seebeck}.
\section{Model Hamiltonian and Formalism}\label{sec2}
\begin{figure*}[t!]
    \centering
    \includegraphics[scale=0.3]{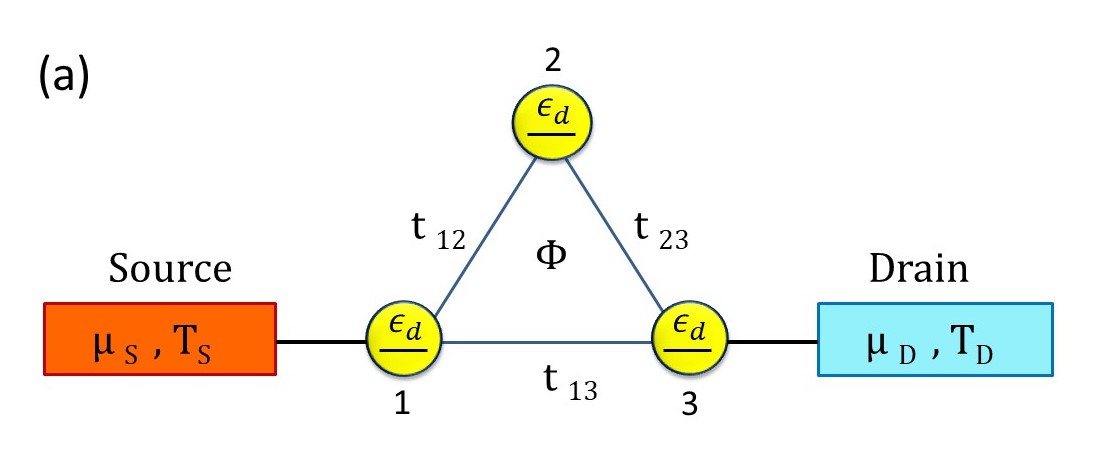}
    \includegraphics[scale=0.3]{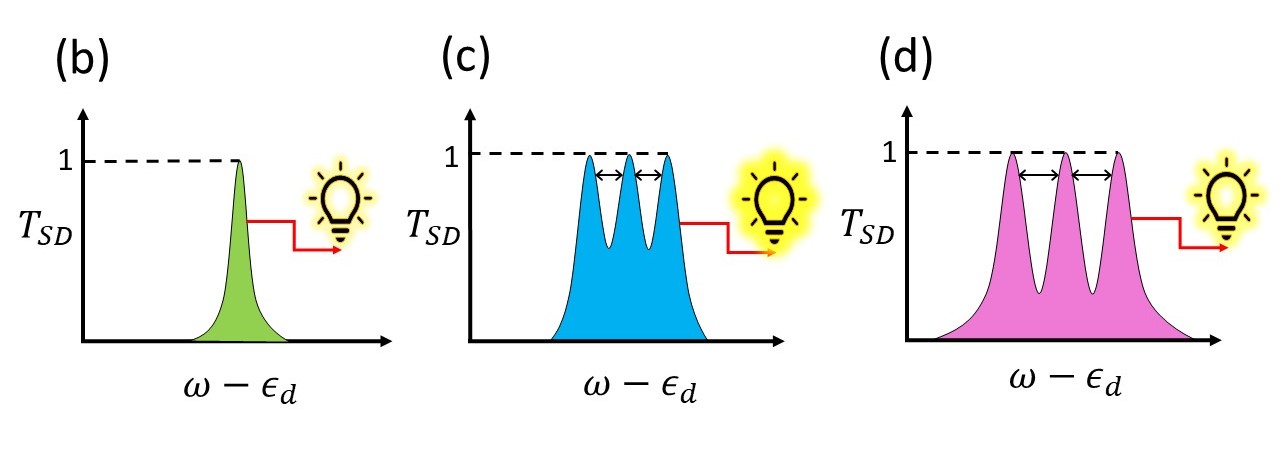}
    \includegraphics[scale=0.3]{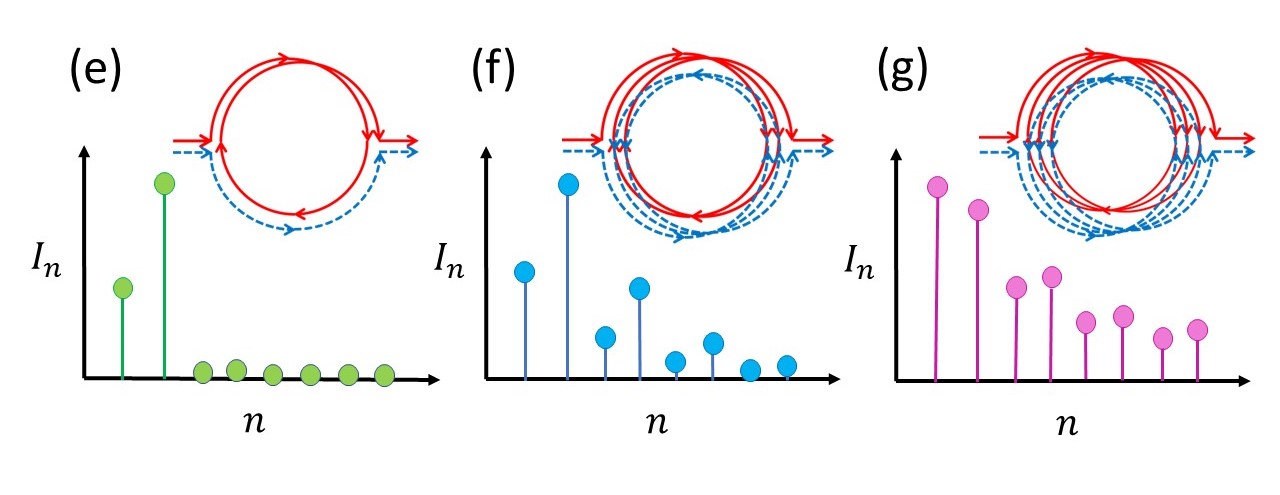}
    \caption{(a) A triple quantum dot Aharonov-Bohm interferometer with dot 1 and dot 3 connected to source and drain, respectively. A magnetic flux $\Phi$ pierces the triangular AB ring perpendicularly. Figures (b)-(d) show transmission patterns with respect to energy for the regimes: $t<\gamma$, $t=\gamma$, and $t>\gamma$, respectively. (b) A single symmetric peak around resonance is more efficient but generates less power ($t<\gamma$). (c) We need three peaks equally spaced around the resonance to achieve optimal power and efficiency as shown in the figure with the brightest bulb ($t=\gamma$). (d) When the separation between the peaks increases antiresonance is more pronounced causing a decrease in power and efficiency ($t>\gamma$). The bottom set of figures shows harmonic patterns of charge current obtained by Fourier decomposition with respect to $\phi$ ($n=1, 2, 3, ...$), and the different trajectories of the circulation of electrons around the AB ring are represented by blue and red orbits. The presence of higher harmonic modes is necessary but not sufficient to achieve optimal power and efficiency. Figure (e) represents only two modes of dominance in the $t<\gamma$ regime responsible for the least output power but highest efficiency; (f) gives optimal power efficiency for the regime $t=\gamma$ with a few higher harmonics; (g) The dominance of higher harmonic modes is responsible for reducing power output as well as the efficiency in the $t>\gamma$ regime. }
    \label{model}
\end{figure*}
We consider a model setup of a triple-quantum-dot Aharanov-Bohm (AB) interferometer \cite{TQD1, TQD2, TQD3, TQD4, BANDYOPADHYAY2021114786}, as shown in Fig. \ref{model}(a). Here, each quantum dot is located at the vertex of an equilateral triangle and interconnected to each other by quantum tunneling. A magnetic flux $\Phi$ pierces the triangular AB ring perpendicularly. Dot 1 and dot 3 are connected to two metallic leads (bath or reservoir) maintained at different temperatures and chemical potentials. Here we do not consider the electron-electron interactions and the spin degrees of freedom to construct a simple solvable setup. Hence we can describe quantum dots by a spin-less electronic level and ignore the Zeeman effect. The total Hamiltonian $\hat{H}$ of the whole system is given by,
\begin{equation}\label{eq1}
    \hat{H}=\hat{H}_{TQD}+\hat{H}_B+\hat{H}_{TQD,B},
\end{equation}
where $\hat{H}_{TQD}$ is the Hamiltonian for the three quantum dots (subsystem), $\hat{H}_B$ is the Hamiltonian for the metallic leads (bath), and $\hat{H}_{TQD, B}$ is the interaction Hamiltonian between the subsystem and the baths. The subsystem Hamiltonian consisting of three interconnected dots is given as
\begin{equation}\label{eq2}
\hat{H}_{TQD}=\sum_{i=1,2,3}\epsilon_i\hat{d}_i^{\dagger}\hat{d}_i+
    \Big[\sum_{i\ne j}t_{ij}\hat{d}_i^{\dagger}\hat{d}_je^{i\phi_{ij}}+\textrm{h.c.}\Big].
\end{equation}
Here, $\epsilon_i$ denotes the energy of the $i^{th}$ dot. $\hat{d}_i^{\dagger}$ and $\hat{d}_i$ are the creation and annihilation operators of the electrons in the respective dots and $t_{ij}$ is the inter-dot tunneling strength and $\phi_{ij}$ is the AB phase factor. The Hamiltonian for the two metallic leads, source (S) and drain (D), consisting of noninteracting electrons is
\begin{equation}\label{eq3}
    \hat{H}_B=\sum_{k,\nu\in{S,D}}\epsilon_{\nu,k}\hat{c}_{\nu,k}^{\dagger}
    \hat{c}_{\nu,k},
\end{equation}
where $\hat{c}_{\nu,k}^{\dagger}$ and $\hat{c}_{\nu,k}$ represents the creation and annihilation operators of the electrons in the $k^{th}$ momentum state and $\epsilon_{\nu,k}$ is the energy of the $k^{th}$ state in the corresponding baths, $\nu\in{S, D}$. Since dot 1 is connected to the source (S) and dot 3 is connected to the drain (D) (see Fig. \ref{model}(a)), we can write the subsystem-bath interaction term as follows
\begin{equation}\label{eq4}
    \hat{H}_{TQD,B}=V_{1,k}^S\hat{d}_1^{\dagger}\hat{c}_{S,k}+
    V_{3,k}^D\hat{d}_3^{\dagger}\hat{c}_{D,k}+\textrm{h.c.}.
\end{equation}
Here $V_{i,k}^{\nu}$ denotes the dot-bath coupling strength. The AB phases $\phi_{ij}$ satisfy the following relation \cite{BANDYOPADHYAY2021114786},
\begin{equation}\label{eq5}
    \phi_{12}+\phi_{23}+\phi_{31}=\phi=2\pi\frac{\Phi}{\Phi_0},
\end{equation}
where $\Phi$ is the total magnetic flux enclosed by the triangular AB ring and $\Phi_0=h/e$ is the flux quantum. In the steady state, physical observables are gauge invariant. Since the dots are present at the vertices of an equilateral triangular loop, we may choose the gauge as $\phi_{12}=\phi_{23}=\phi_{31}=\phi/3$. For our system, we maintained a symmetric voltage bias condition i.e., $\mu_S=-\mu_D$. Although, by applying a gate voltage to each dot we can place the levels of the dots away from the symmetric point at which $\mu_S-\epsilon_i=\epsilon_i-\mu_D$. For simplicity, we use the natural unit convention $\hbar=c=e=k_B=1$. We translate our results to physical units in Appendix \ref{appendixB}.
\subsection{Non-equilibrium Green's functions for the AB interferometer setup}
To solve this model and compute the observables of interest, we use the Nonequilibrium Green's Function (NEGF) approach \cite{meir1992, wangNEGF, textbook}. We follow the equation of motion method for the calculations (for details see Appendix \ref{appendixA}) and obtain the retarded [$G^+(\omega)$] and advanced [$G^-(\omega)$] Green's function for our system by using the quantum Langevin equation \cite{Dhar2006}. The Green's functions are given as,
\begin{equation}\label{eq12}
    G^{\pm}(\omega)=\Big[\omega I-H_{TQD}-
    \Sigma_S^{\pm}(\omega)-\Sigma_D^{\pm}(\omega)\Big]^{-1},
\end{equation}
where $I$ is a $(3\times3)$ identity matrix. Here $\Sigma_S^{\pm}(\omega)$  and $\Sigma_D^{\pm}(\omega)$ are the self-energies defined in Eq. (\ref{eq13}) of Appendix \ref{appendixA}. $H_{TQD}$ is the single particle matrix corresponding to the Hamiltonian $\hat{H}_{TQD}$ in Eq. (\ref{eq2}) and given by,
\begin{equation}
\renewcommand{\arraystretch}{2}
    H_{TQD}=
    \begin{pmatrix}
    \epsilon_d & te^{i\phi/3} & te^{-i\phi/3}\\
    te^{-i\phi/3} & \epsilon_d & te^{i\phi/3}\\
    te^{i\phi/3} & te^{-i\phi/3} & \epsilon_d
    \end{pmatrix}.
\end{equation}
Note that, we impose energy degeneracy for the dots $\epsilon_1=\epsilon_2=\epsilon_3=\epsilon_d$ and further consider identical symmetric inter-dot tunneling strength as $t_{12}= t_{23}= t_{13}=t$.  In the wide-band limit when the density of states of the metallic lead is energy independent, the real part of the self-energy term vanishes. Then we can define the hybridization matrix from the relation $\Sigma^{+}=-i\Gamma/2$:
\begin{equation}\label{eq14}
    \Gamma_{i,i'}^{\nu}=2\pi\sum_{k,\nu}V_{i',k}^{\nu^*}V_{i,k}^{\nu}\, \delta(\omega-\omega_k).
\end{equation}
We may take $V_{i,k}^{\nu}$ as real constants with $\nu=S, D$, independent of the level index and reservoir state, resulting in $\Gamma_{1, 1}^S=\gamma_S$ and $\Gamma_{3, 3 }^D=\gamma_D$, where $\gamma_{\nu}$ describes the coupling between the dots and metallic leads and it is taken as a constant (energy independent).\\
We then receive the retarded Green's function as,
\begin{equation}\label{eq15}
\renewcommand{\arraystretch}{2}
    G^{+}(\omega)=
    \begin{pmatrix}
    \omega-\epsilon_d+i\frac{\gamma_S}{2} & -te^{i\phi/3} & -te^{-i\phi/3}\\
    -te^{-i\phi/3} & \omega-\epsilon_d & -te^{i\phi/3}\\
    -te^{i\phi/3} & -te^{-i\phi/3} & \omega-\epsilon_d+i\frac{\gamma_D}{2}
    \end{pmatrix}^{-1}.
\end{equation}
and the hybridization matrices are given by
\begin{equation}\label{eq16}
    \Gamma^S=
    \begin{pmatrix}
    \gamma_S & 0 & 0\\
    0 & 0 & 0\\
    0 & 0 & 0
    \end{pmatrix},
    \Gamma^D=
    \begin{pmatrix}
    0 & 0 & 0\\
    0 & 0 & 0\\
    0 & 0 & \gamma_D
    \end{pmatrix}.
\end{equation}
We can define the advanced Green's function as conjugate transpose matrix $G^{-}(\omega)=[G^+(\omega)]^{\dagger}$. The details are discussed in Appendix \ref{appendixA}.
\subsection{Observables: Currents, Output Power and Efficiency}\label{sec2b}
It is of interest to investigate the subsystem properties and obtain the transmission coefficient, particle currents, and heat currents through the system. The transmission of electrons from reservoir $\nu$ to $\xi$ is given by the transmission coefficient \cite{datta1997electronic}
\begin{equation}\label{eq17}
    T_{\nu\xi}(\omega,\phi)=\mathrm{Tr}\big[\Gamma^{\nu}G^{+}(\omega,\phi)\Gamma^{\xi}G^{-}(\omega,\phi)\big].
\end{equation}

Using the transmission coefficients $T_{\nu\xi}$, we can express the particle currents flowing from reservoir $\nu$ to the central system as \cite{current1, current2}
\begin{equation}\label{eq18}
    I_{\nu}=\int_{-\infty}^{\infty}d\omega\sum_{\xi\ne\nu}\big[
    T_{\nu\xi}(\omega,\phi)f_{\nu}(\omega)-
    T_{\xi\nu}(\omega,\phi)f_{\xi}(\omega)\big].
\end{equation}
\indent
Although, the definition of the heat current is debatable \cite{hanggi} in the strong coupling regime, we consider the conventional and usefully studied nonequilibrium Green's function approach to define the heat current from the reservoir $\nu$ for arbitrary coupling as \cite{casati,stafford,galperin2,galperin3, galperin4,galperin1,topp}
\begin{equation}\label{eq19}
    \begin{split}
       Q_{\nu}=\int_{-\infty}^{\infty}d\omega(\omega-\mu_{\nu})
        \sum_{\xi\ne\nu}\big[T_{\nu\xi}(\omega,\phi)f_{\nu}(\omega) \\
         -T_{\xi\nu}(\omega,\phi)f_{\xi}(\omega)\big],
    \end{split}
\end{equation}
where $f_{\nu(\xi)}(\omega)=[e^{(\omega-\mu_{\nu(\xi)})/T_{\nu(\xi)}}+1]^{-1}$ is Fermi distribution function of the reservoir $\nu(\xi)=S,D$ with $\mu_{\nu}$ and $T_{\nu}$ be the corresponding chemical potential and temperature, respectively.\\
\indent To characterize the nonlinear thermoelectric performance of a two-terminal system as a heat engine, we shall study the output power $P$ and the steady-state heat-to-work conversion efficiency $\eta$. Our model uses the configuration $\Delta\mu=\mu_D-\mu_S>0$ and $\Delta T=T_S-T_D>0$. The output power $P$ is equal to the sum of all the heat currents exchanged between the subsystem and the reservoir and it is given by \cite{casati, mazza2014thermoelectric}
\begin{equation}\label{eq20}
    P=\sum_{\nu=S,D}Q_{\nu}=(\mu_D-\mu_S)I_S.
\end{equation}
Equation (\ref{eq20}) follows the laws of conservation of particle $\sum_{\nu}I_{\nu}=0$ and energy. We define the efficiency $\eta$ as the ratio of output power $P$ to the heat currents absorbed from the hot bath and it is expressed as \cite{casati, mazza2014thermoelectric}
\begin{equation}\label{eq21}
    \eta=\frac{P}{Q_S}.
\end{equation}
The system works as a heat engine for positive output power $P>0$ with positive heat current flow from the source $Q_S>0$. The efficiency $\eta$ is bounded from the above by Carnot efficiency $\eta_C=1-T_c/T_h$ with $T_c$ and $T_h$ being the temperatures of cold and hot baths, respectively \cite{carnot1824reflections}.
\section{Transport properties: Density of States and Transmission function}\label{occup}
\begin{figure}[t!]
    \centering
    \includegraphics[width=4.27cm]{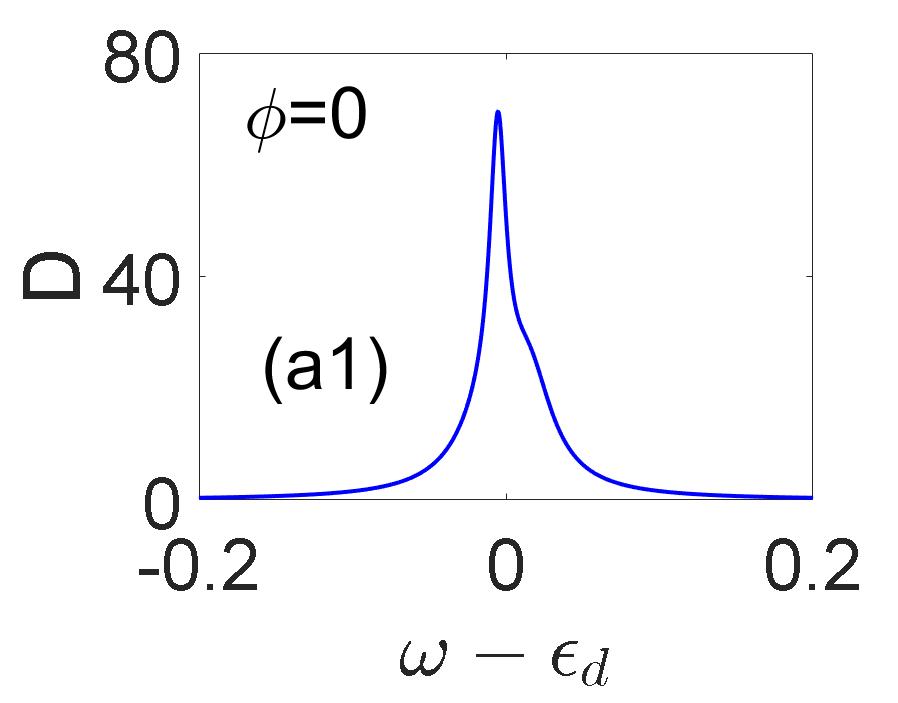}
    \includegraphics[width=4.27cm]{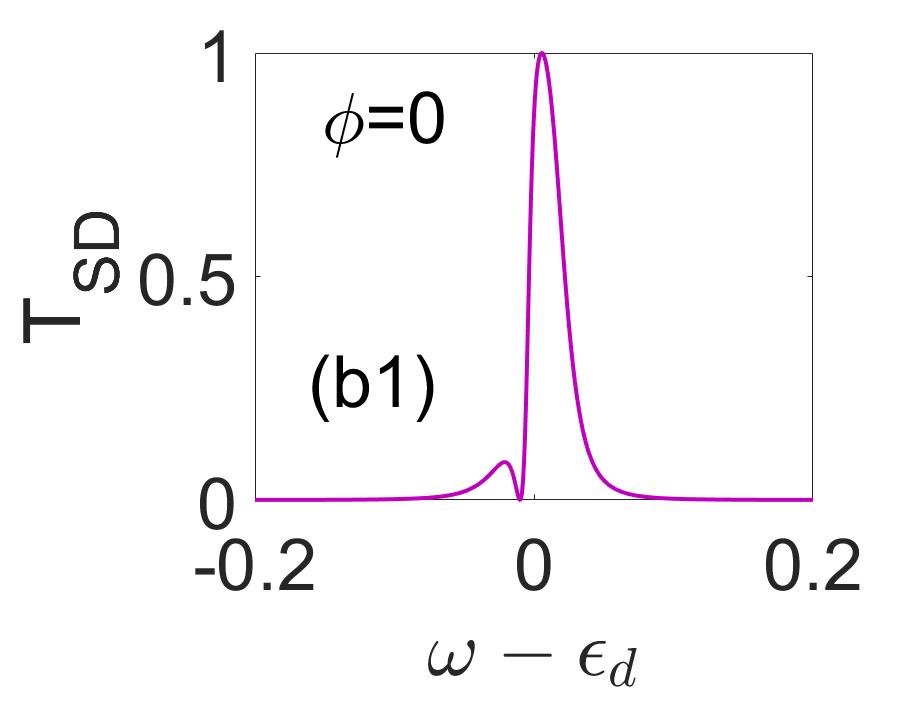}
    \includegraphics[width=4.27cm]{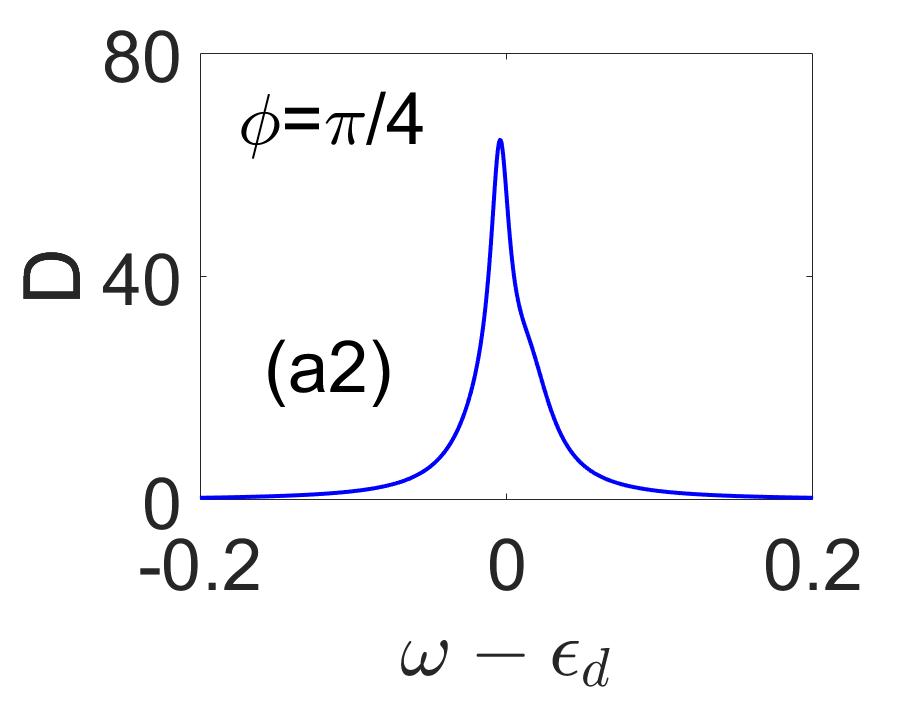}
    \includegraphics[width=4.27cm]{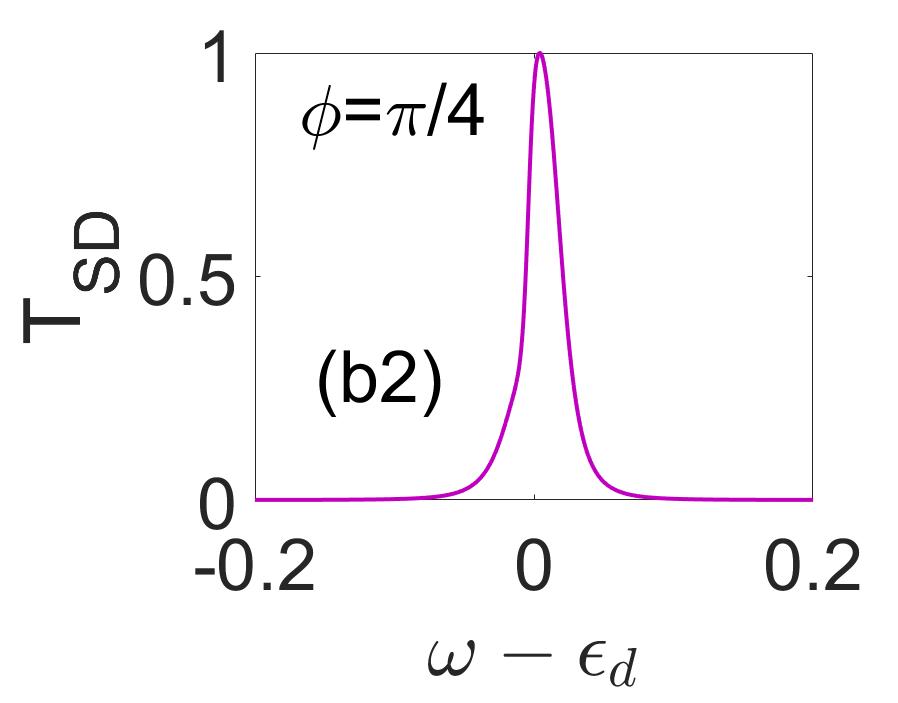}
    \includegraphics[width=4.27cm]{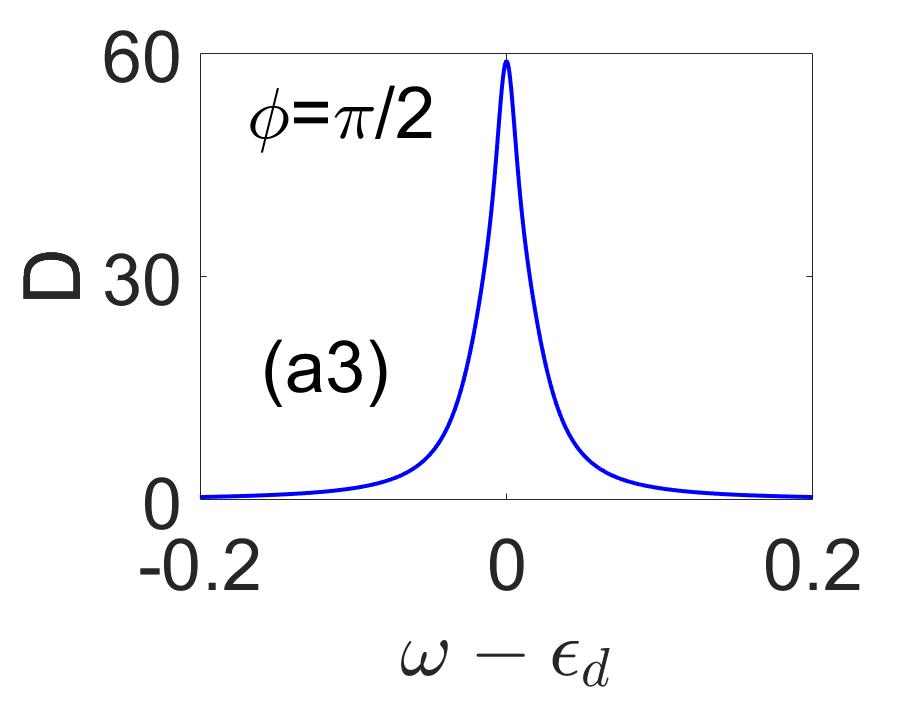}
    \includegraphics[width=4.27cm]{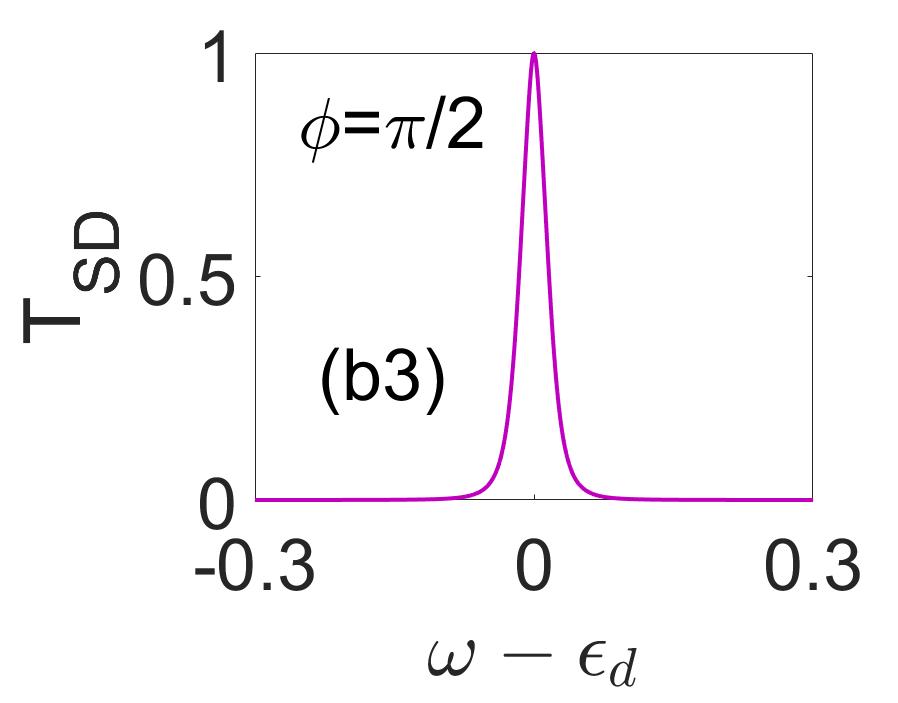}
    \caption{Density of states (a1-a3) and transmission functions (b1-b3) as a function of energy at different values of $\phi$ for $t<\gamma$ regime. Parameters used are $\gamma=0.05$, $t=0.2\gamma$, $\epsilon_d=8\gamma$, $T_S=12\gamma$, $T_D=2\gamma$, $\mu_S=-4\gamma$, $\mu_D=4\gamma$.}
    \label{trans_less}
\end{figure}

The main objective of our work is to find the optimal power-efficiency configuration of a triple-dot heat engine. Intuitively, we expect the maximal constructive interference arising from multiple paths winding around the loop to enhance the power-efficiency. Since these interference properties are encoded in the systems transmission function, we investigate its behavior in three different regimes: (i) $t/\gamma<1$ (ii) $t/\gamma=1$ and (iii) $t/\gamma>1$. We further find that the behavior of the transmission function and the density of states are identical. This provides a prescription for choosing optimal parameters.\\
\indent
The density of states (DOS) is given by the trace of the imaginary part of the retarded Green's function as \cite{datta1997electronic}
\begin{equation}
    D(\omega,\phi)=-\frac{1}{\pi}\mathrm{Tr}[\mathrm{Im}(G^+)],
\end{equation}
which for our setup is given as,
\begin{equation}
    \begin{split}
        D(\omega,\phi)=\frac{\gamma}{\pi\Delta(\omega,\phi)}\Big[(\omega-\epsilon_d)^4+\frac{\gamma^2}{4}(\omega-\epsilon_d)^2 \\
        + 4t^3\cos{\phi}(\omega-\epsilon_d)+t^2(3t^2+\frac{\gamma^2}{4})\Big],
    \end{split}
\end{equation}
where
\begin{equation}
    \begin{split}
        \Delta(\omega,\phi)=\Big[(\omega-\epsilon_d)\Big((\omega-\epsilon_d)^2-3t^2-\frac{\gamma^2}{4}\Big)-2t^3\cos{\phi}\Big]^2 \\
        +\gamma^2\Big[(\omega-\epsilon_d)^2-t^2\Big]^2.
    \end{split}
\end{equation}
We can express the transmission function from reservoir $S$ to $D$ with symmetric dot-lead coupling ($\gamma_S=\gamma_D=\gamma$) analytically as
\begin{equation}\label{trans}
    T_{SD}(\omega, \phi)=\frac{\gamma^2\big[t^4+2t^3\cos{\phi}(\omega-\epsilon_d)+t^2(\omega-\epsilon_d)^2\big]}{\Delta(\omega,\phi)}.
\end{equation}
For $\phi=\pi/2$, the transmission peaks [$T_{SD}(\omega,\phi)=1$] are at positions
\begin{equation}\label{peaks}
    \omega=\epsilon_d, \hspace{0.5cm} \textrm{and} \quad \quad  \omega=\epsilon_d\pm\frac{1}{2}\sqrt{12t^2-\gamma^2}.
\end{equation}
We have three transmission peaks at energies given by Eq. (\ref{peaks}) for all the three regimes i.e., (i) $t<\gamma$, (ii) $t=\gamma$ and (iii) $t>\gamma$. But, when $t\ll\gamma$, we get a single transmission peak at $\omega=\epsilon_d$.\\
\indent
Considering our model set-up where the triple-dot AB interferometer is connected to two reservoirs or leads, the transmission function describes the quantum interference. The transmission function can exhibit both constructive and destructive interference. Further, we observe that DOS is directly related to the transmission function.
We analyze this quantity for the three different regimes. In Fig. \ref{trans_less}(a1), for $t<\gamma$, we observe an asymmetric structure in DOS with energy when the magnetic flux is zero i.e., $\phi=0$. As we change the flux to $\phi=\pi/4$ and $\pi/2$, the asymmetry vanishes and we get a single peak (see Fig. \ref{trans_less}(a2), (a3)). We can also see this behavior in the transmission function. We observe an asymmetric antiresonance dip in transmission at $\phi=0$ as shown in Fig. \ref{trans_less}(b1) and as we increase the magnetic flux to $\phi=\pi/2$ the asymmetry in transmission vanishes and we get a single peak symmetric around $\omega=\epsilon_d$ (see Fig. \ref{trans_less}(b3)).
It has also been observed that the DOS and transmission function behavior for $\phi=\pi$, $\phi=3\pi/4$ has reflection symmetry of that of $\phi=0$ and $\phi=\pi/4$, respectively. This is also true for other values of $\phi$.\\
\begin{figure}[t]
    \centering
    \includegraphics[width=4.27cm]{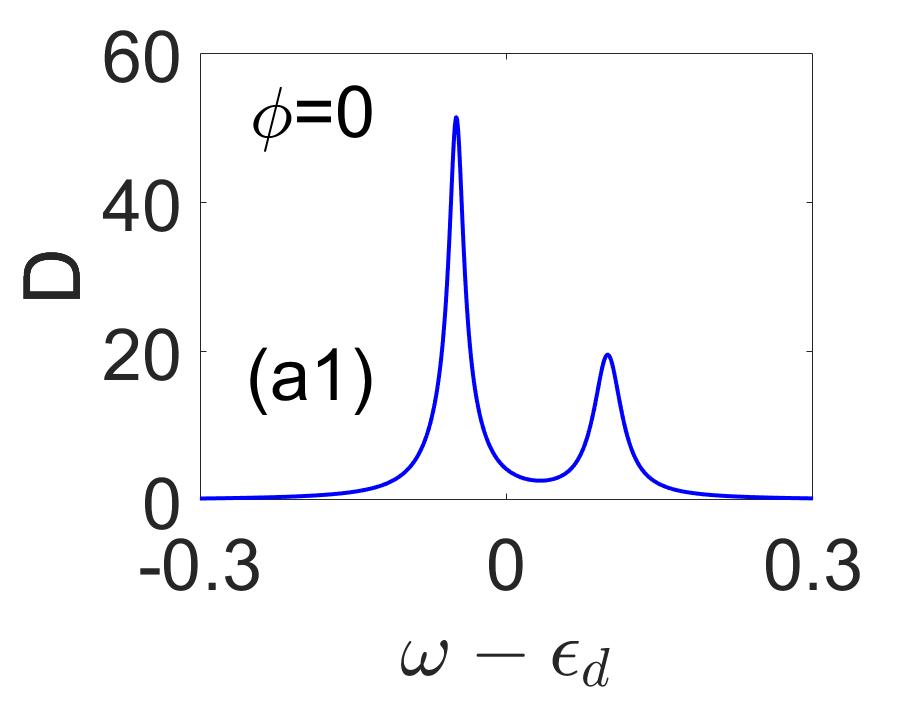}
    \includegraphics[width=4.27cm]{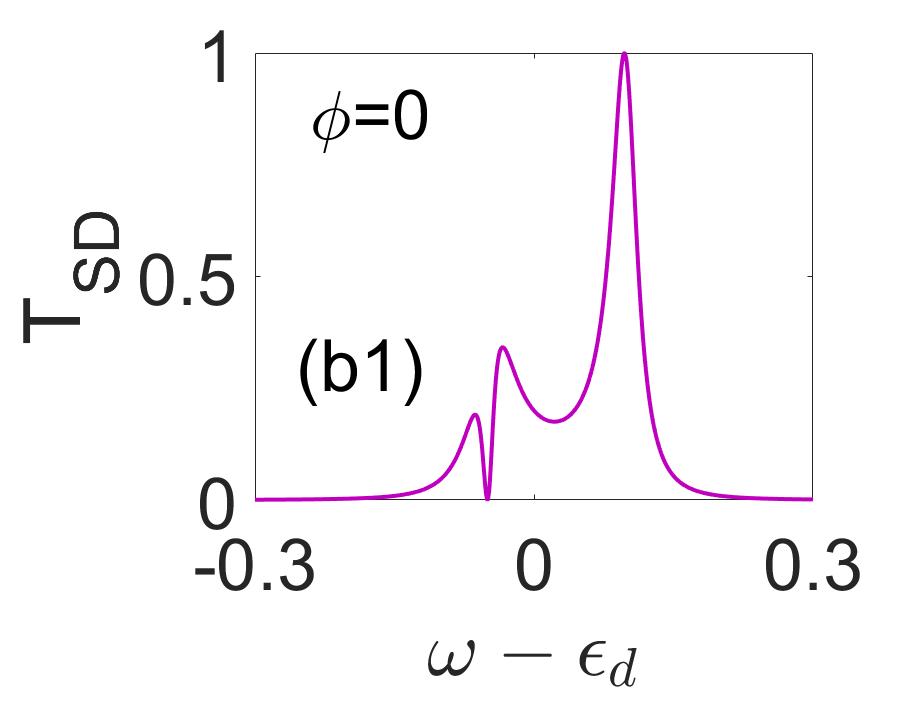}
    \includegraphics[width=4.27cm]{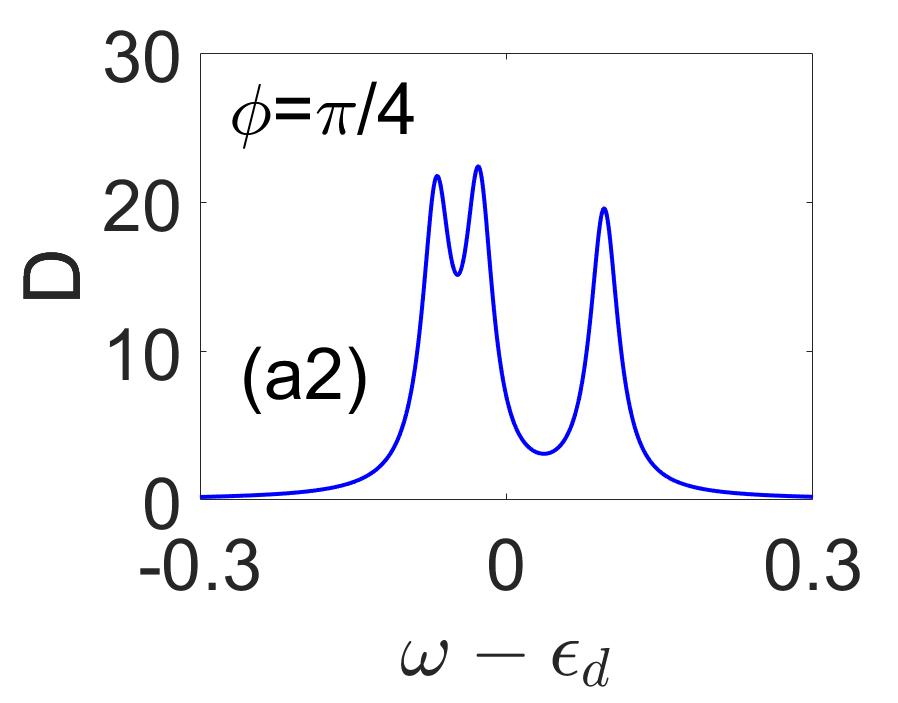}
    \includegraphics[width=4.27cm]{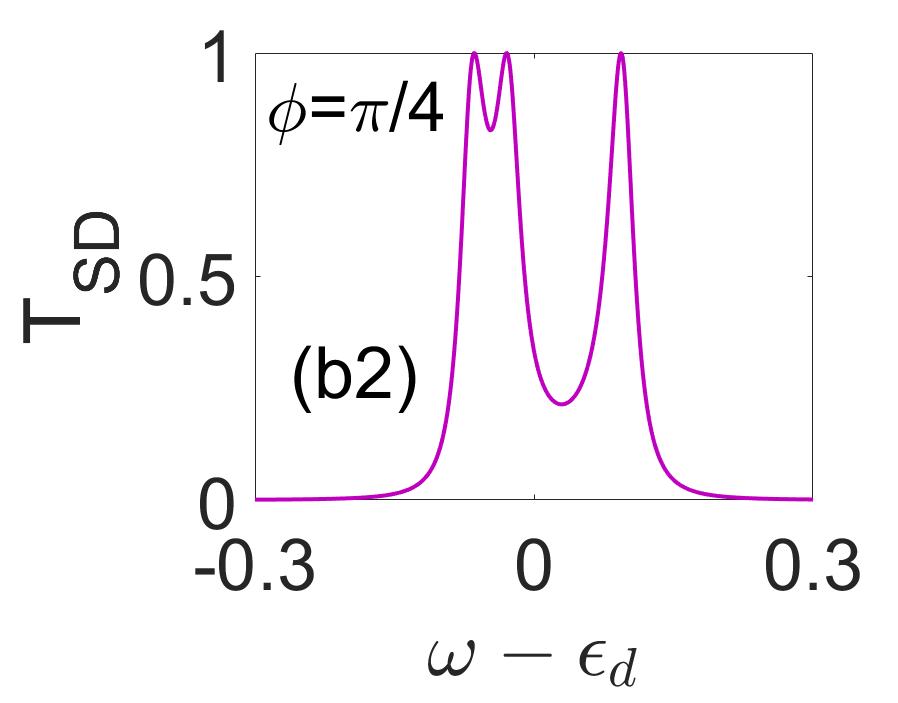}
    \includegraphics[width=4.27cm]{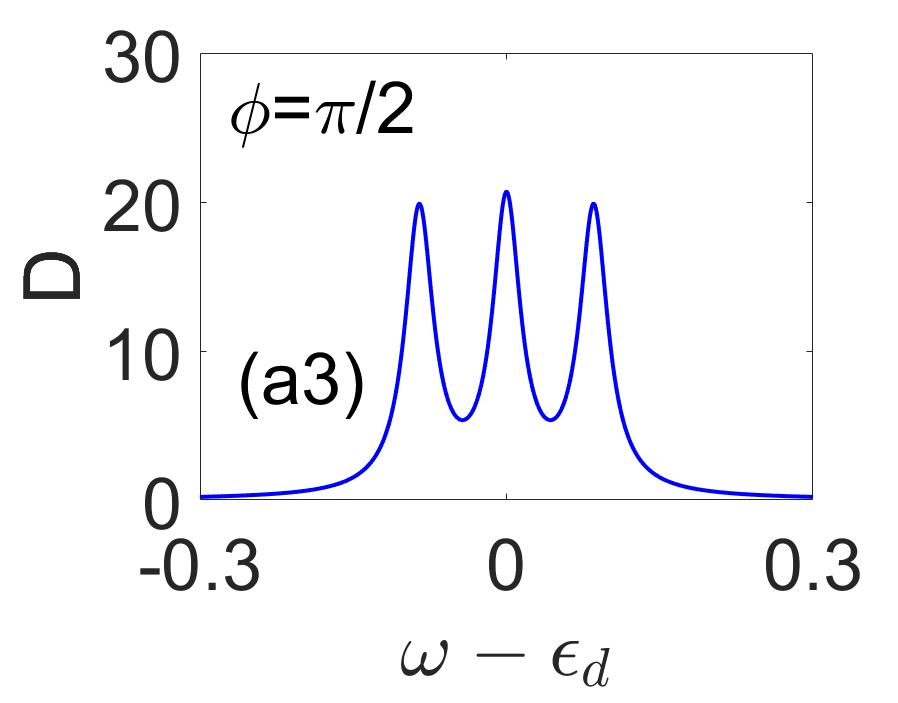}
    \includegraphics[width=4.27cm]{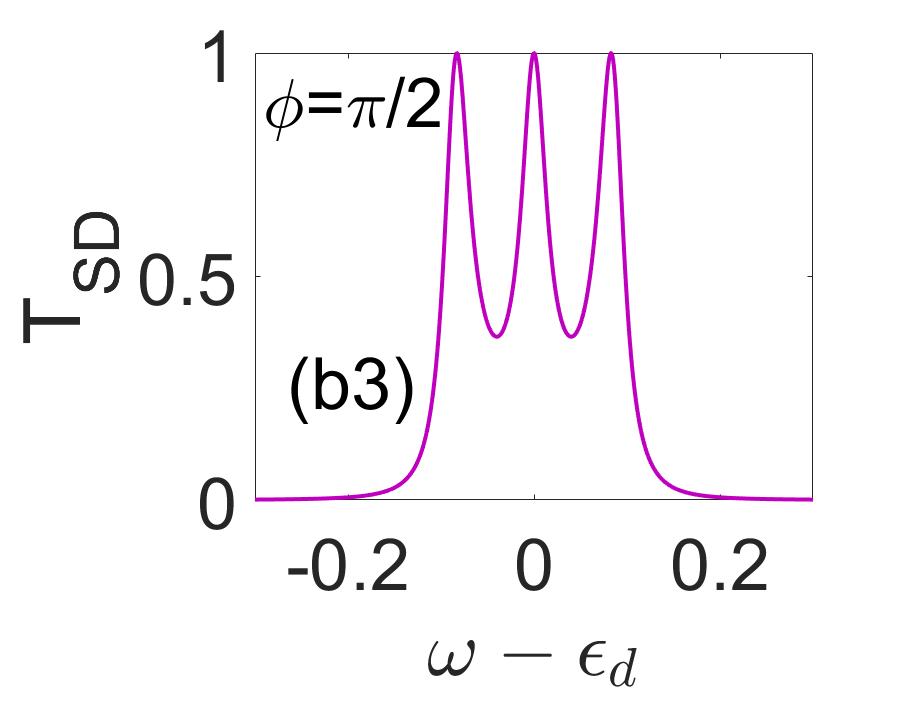}
    \caption{Density of states (a1-a3) and transmission functions (b1-b3) as a function of energy at different values of $\phi$ for $t=\gamma$ regime. Parameters used are $\gamma=0.05$, $t=\gamma$, $\epsilon_d=8\gamma$, $T_S=12\gamma$, $T_D=2\gamma$, $\mu_S=-4\gamma$, $\mu_D=4\gamma$.}
    \label{trans_equal}
\end{figure}
\indent
In the $t\sim \gamma$ regime, we can observe that the DOS has two peaks and the transmission has antiresonance dip for $\phi=0$ in Fig. \ref{trans_equal}(b1). Further, as we change the magnetic flux from $\phi=0$ to $\phi=\pi/4$ the DOS and the transmission function exhibit three resonance peaks. For $\phi=\pi/2$, resonance peaks become sharper and are symmetrically situated around $\omega=\epsilon_d$ (see Fig. \ref{trans_equal}(b3)) such that the maximum transmission is one and the minimum is half. It opens up more transport channels and all three states participate equally in transport for $\phi=\pi/2$, $t=\gamma$ leading to perfectly constructive interference.
\begin{figure}[t]
    \centering
    \includegraphics[width=4.27cm]{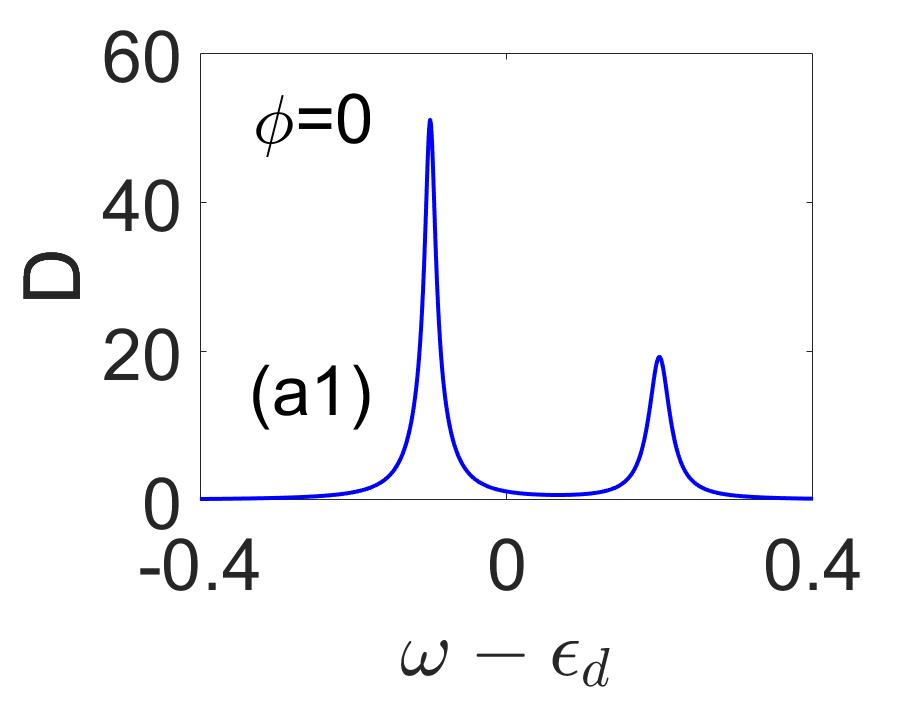}
    \includegraphics[width=4.27cm]{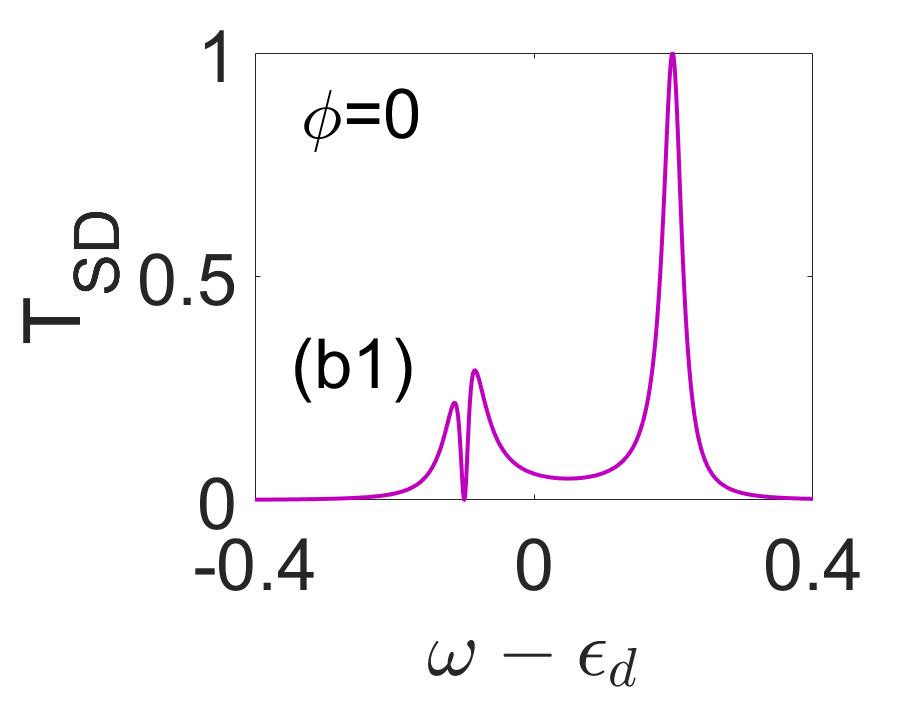}
    \includegraphics[width=4.27cm]{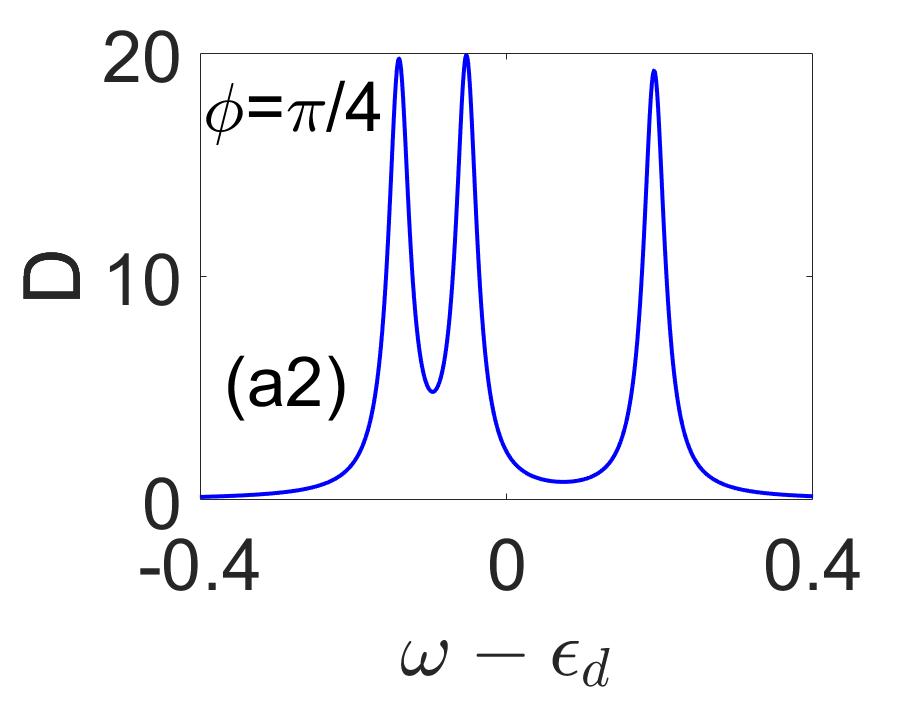}
    \includegraphics[width=4.27cm]{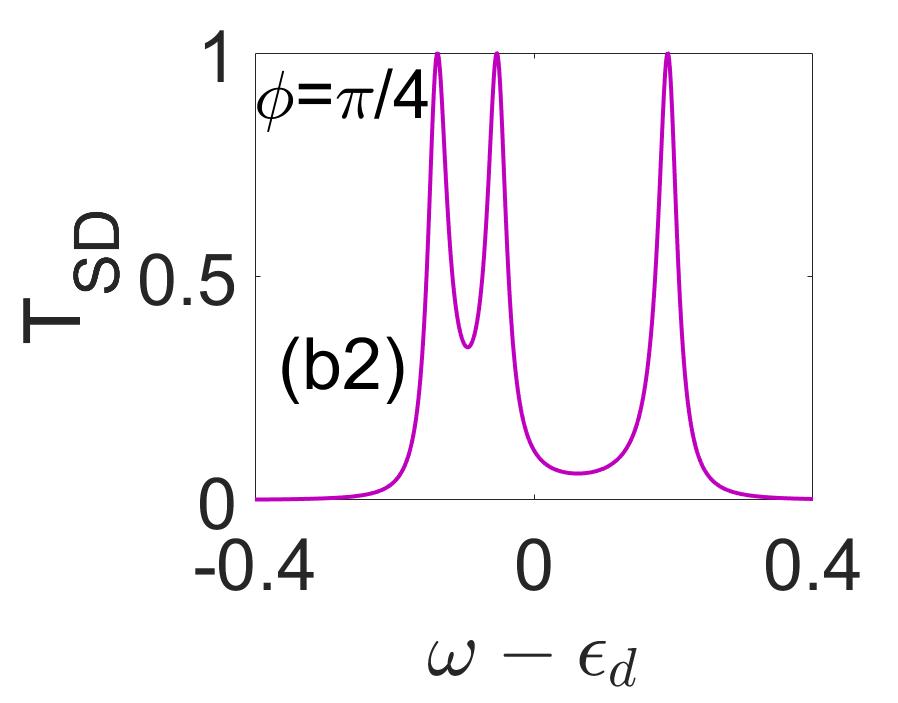}
    \includegraphics[width=4.27cm]{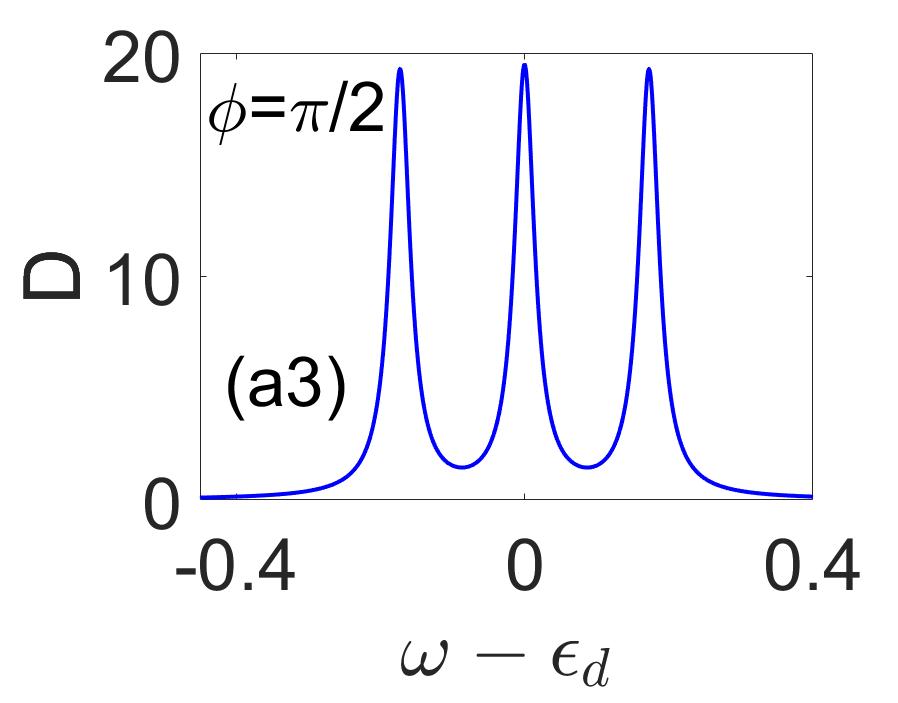}
    \includegraphics[width=4.27cm]{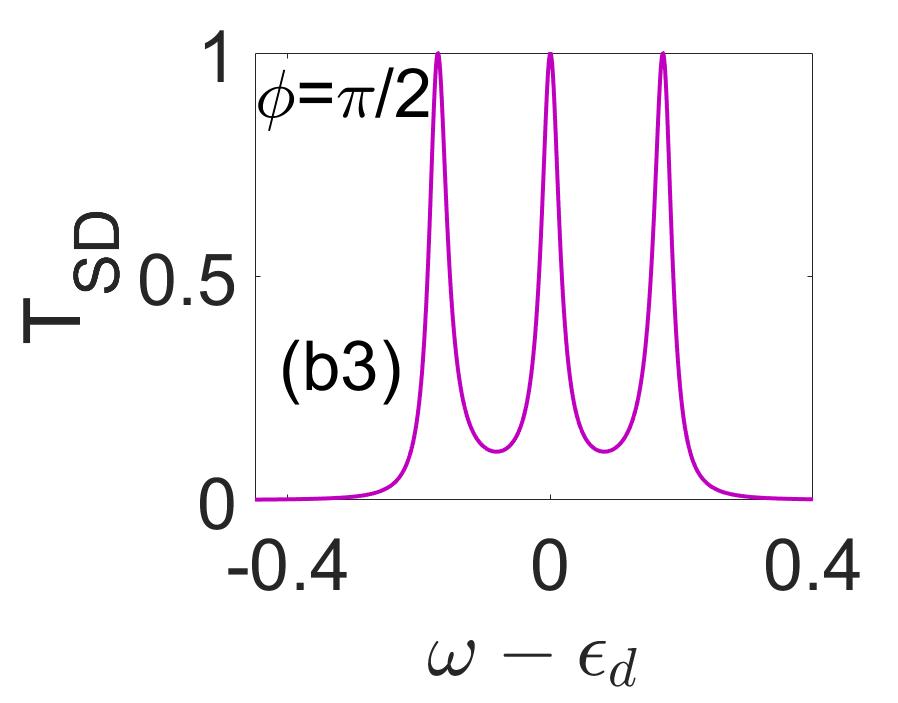}
    \caption{Density of states (a1-a3) and transmission functions (b1-b3) as a function of energy at different values of $\phi$ for $t>\gamma$ regime. Parameters used are $\gamma=0.05$, $t=0.2\gamma$, $\epsilon_d=8\gamma$, $T_S=12\gamma$, $T_D=2\gamma$, $\mu_S=-4\gamma$, $\mu_D=4\gamma$.}
    \label{trans_large}
\end{figure}
For the $t>\gamma$ regime as shown in Fig. \ref{trans_large}, we observe that the DOS and transmission function show a double resonance peak for $\phi=0$. As one increases $\phi$, we observe the splitting of the peaks of the DOS and transmission function. One can observe three equally spaced resonance peaks for $\phi=\pi/2$ symmetric around $\epsilon_d$. The separation between the resonance peaks becomes wider for both DOS and the transmission function and it is accompanied by two deep antiresonance dips as compared to the $t=\gamma$ regime.\\
\indent
 From the above analysis, we can conclude that the behavior of the transmission function and the density of states are identical. This suggests a peculiar way of engineering quantum-dot structures to achieve the maximal constructive interference condition.
 We can explain the antiresonance dip in transmission i.e., $T_{SD}=0$ for $\phi=0$ in all the three regimes from the energy spectrum of the isolated triple-dot AB setup. The strong (compared to lead-dot coupling) inter-dot tunneling coupling leads to a strong hybridization between triple-dot states. Owing to the discrete molecular symmetry of the setup, the spectrum exhibits degeneracies depending on the magnetic flux values. In the case of an isolated symmetric triple dot, it can be shown that the energy eigenstates are Bloch
waves uniformly delocalized over three sites. When a pair
of degenerate levels is in resonance with incoming electrons
from leads, the transmission probability vanishes which
is a consequence of destructive interference. Further, if the contact
with the lead is weak, then the energy-level structure of
the isolated triple-dot will be reflected in the plot of the transmission function, and $T_{SD}$ will
possess a series of peaks depending on the energy of the incoming electron located at energies in the vicinity of the energy spectrum. We observe this behavior from $t\sim\gamma$ and $t>\gamma$ regimes from Fig. \ref{trans_equal} and Fig. \ref{trans_large}, respectively.
But the separation between the resonance peaks becomes wider and anti-resonance dips become broader as we move from $t\sim\gamma$ to $t>\gamma$.  We observe a perfectly constructive interference for $\phi=\pi/2$, $t\sim\gamma$ and each state contributes equally to the transport. This is analogous to an almost loss-free beam splitter where the transmission through each channel is unity. This would facilitate efficient charge and energy transport. To achieve optimal power efficiency, the maximum transmission should be one and the minimum should be half. The presence of antiresonance dips reduces efficiency. In the subsequent section, we observe that the position of these transmission peaks and their separation are crucial for the power-efficiency trade-off for a thermoelectric heat engine. We also study the efficiency of the heat engine and connect it with multiple-harmonics patterns which can be extracted from the Fourier analysis of the steady-state charge and heat currents.\\
\indent
In summary, we have investigated the transmission function and DOS in order to investigate the relationship
between the line shape of the transmission spectra and the basic electronic structure of our phase-tunable AB heat engine toy model. The QI features in the transmission spectra such as Fano and anti-resonance could be realized and modulated by varying the ratio $\frac{t}{\gamma}$ and external flux $\phi$. This information will be further utilized in the next section to optimize the thermoelectric properties of the heat engine.
\section{Power-efficiency trade-off and analysis of Harmonics in currents}\label{power_harmonics}
\begin{figure}[t]
    \centering
    \includegraphics[width=6cm]{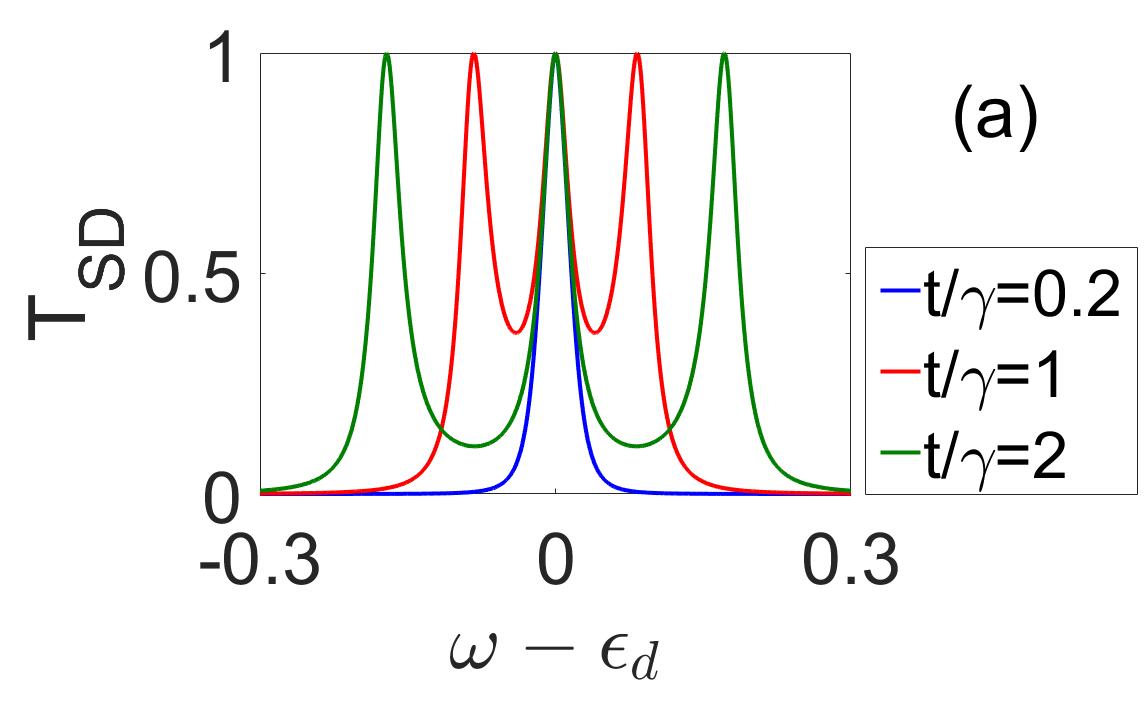}
    \includegraphics[width=6cm]{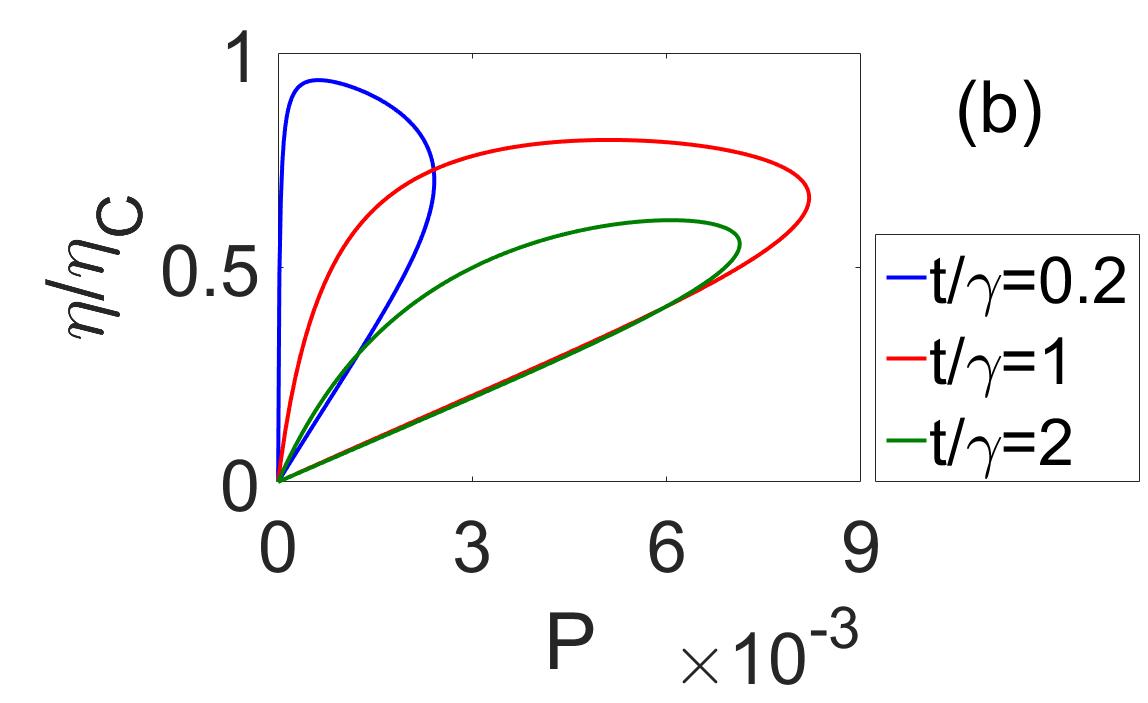}
    \caption{(a) Transmission function and (b) Power-efficiency diagram for different values of tunneling strength $t$. Parameters used are $\gamma=0.05$, $\epsilon_d=8\gamma$, $\mu_S=-\mu_D$, $T_S=12\gamma$, $T_D=2\gamma$, $\phi=\pi/2$.}
    \label{power_effc_t}
\end{figure}
In this section, we analyze the power versus efficiency trade-off in the nonlinear regime as a function of the magnetic flux $\phi$ and the ratio of the intra-dot tunneling rate $t$ and the hybridization strength $\gamma$ i.e., $t/\gamma$. Considering a particular temperature bias, we find the optimal flux and ratio $t/\gamma$ to maximize both output power and efficiency as defined in Eqs. (\ref{eq20}) and (\ref{eq21}), respectively.
We demonstrate the Lasso-type parametric plot of efficiency versus
power and verify the trade-off between both quantities.
We demonstrate that the AB heat engine can be tuned through external
magnetic flux $\phi$, gate voltage $V_g$ (to vary the dot energy), temperature bias $\Delta T$, and the ratio $t/\gamma$  to optimize either efficiency or power or the both.\\
\indent
From Fig. \ref{power_effc_t}(b), it is observed that we get the maximum efficiency for $t<\gamma$, but at the cost of low power output. In this case, we can see that the transmission has less area under the curve, and a single channel participates in the transport. For the $t=\gamma$ regime, we get maximum output power with a larger efficiency. For this case, we find that three resonances are closer and equally contribute. In this regime, irrespective of $\phi$, we find that there are three peaks (not necessarily equally spaced) in the transmission, and we observe the higher harmonics. The magnitude of these harmonics also increases as we move from $\gamma>t$ to $\gamma=t$. We can conclude that $t\sim\gamma$ and $\phi=\pi/2$ is the optimal regime for transport where the transmission function has three peaks with equal separation as shown in Fig. \ref{power_effc_t}(a). Three peak structure of the transmission signals the higher-order interference pattern arising from trajectories winding multiple times within the triple-dot subspace.
There is only one transport channel for $t<\gamma$ and the transmission peaks start separating as we move from the $t<\gamma$ regime towards the $t\sim\gamma$ regime. The separation between the three resonance peaks becomes wider, and the antiresonance is more pronounced as we move from $t\sim\gamma$ to $t>\gamma$ regime. In the $t>\gamma$ regime, inter-dot tunneling is very large compared to the dot-lead hybridization strength. Thus, the electrons that are coming from the source will prefer to circulate in the loop and hence efficiency decreases although power output is still much larger than the $\gamma>t$ case. \\
\indent
\begin{figure}[t!]
    \centering
    \includegraphics[width=4.27cm]{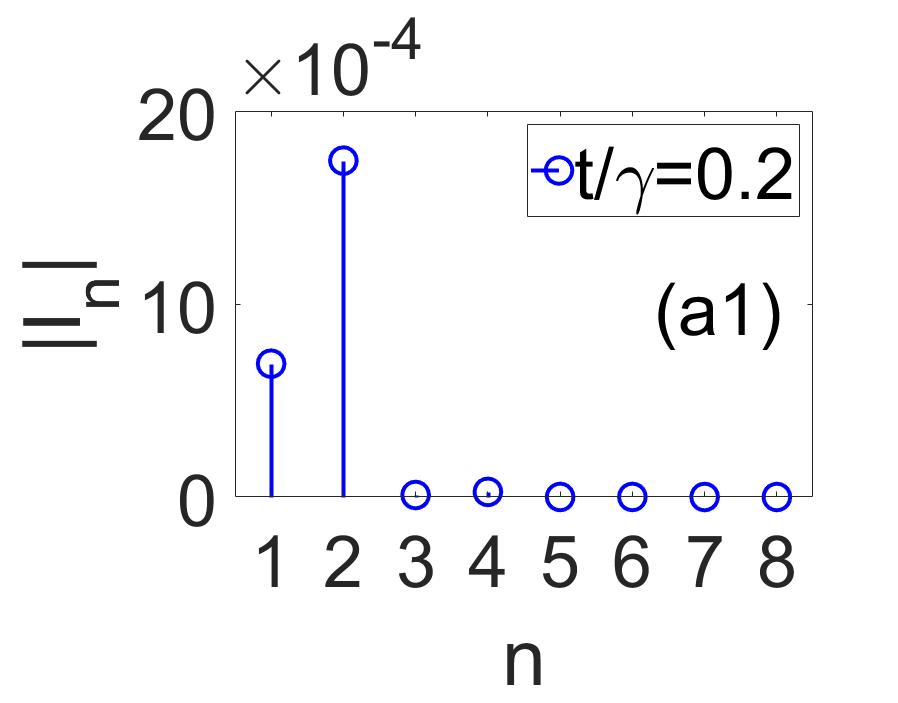}
    \includegraphics[width=4.27cm]{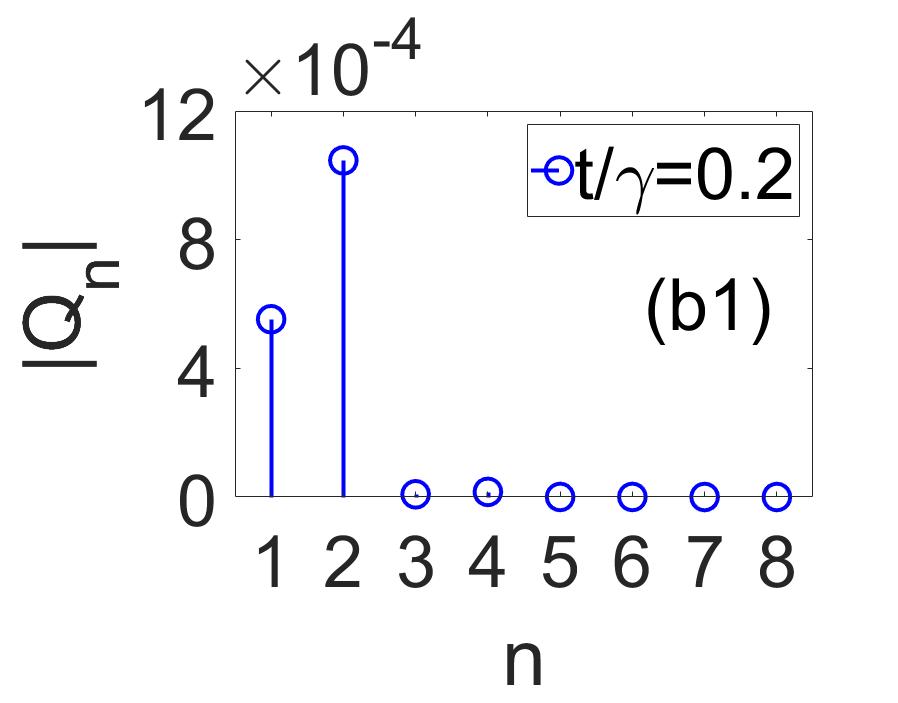}
    \includegraphics[width=4.27cm]{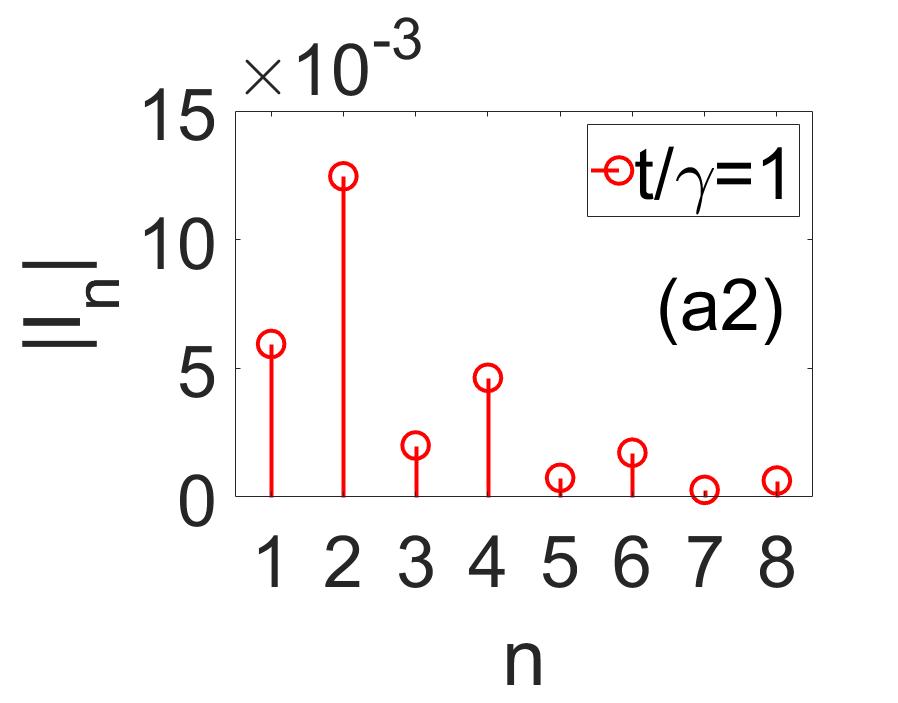}
    \includegraphics[width=4.27cm]{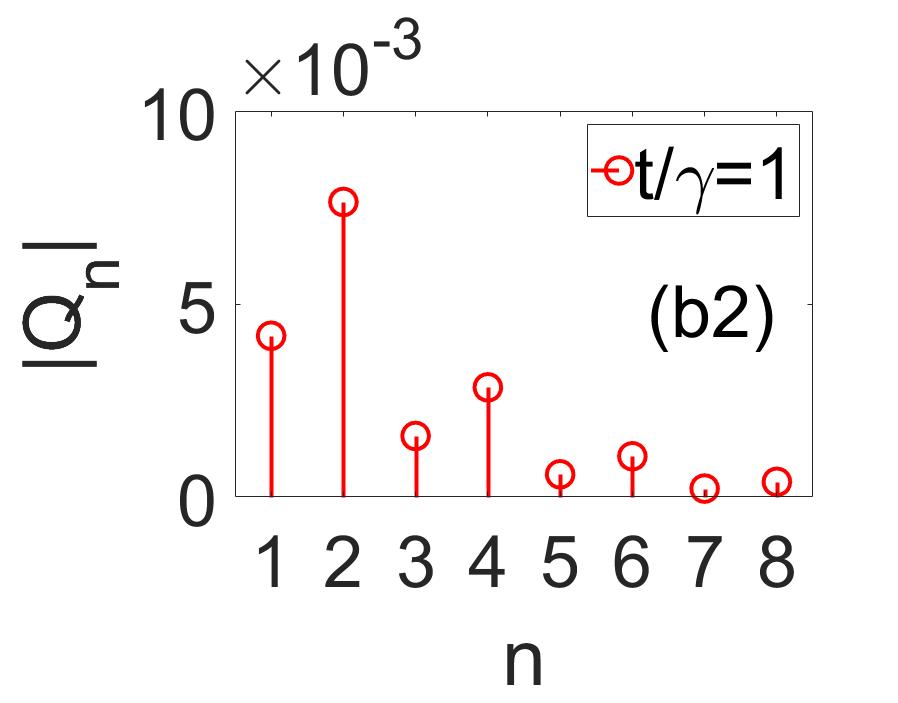}
    \includegraphics[width=4.27cm]{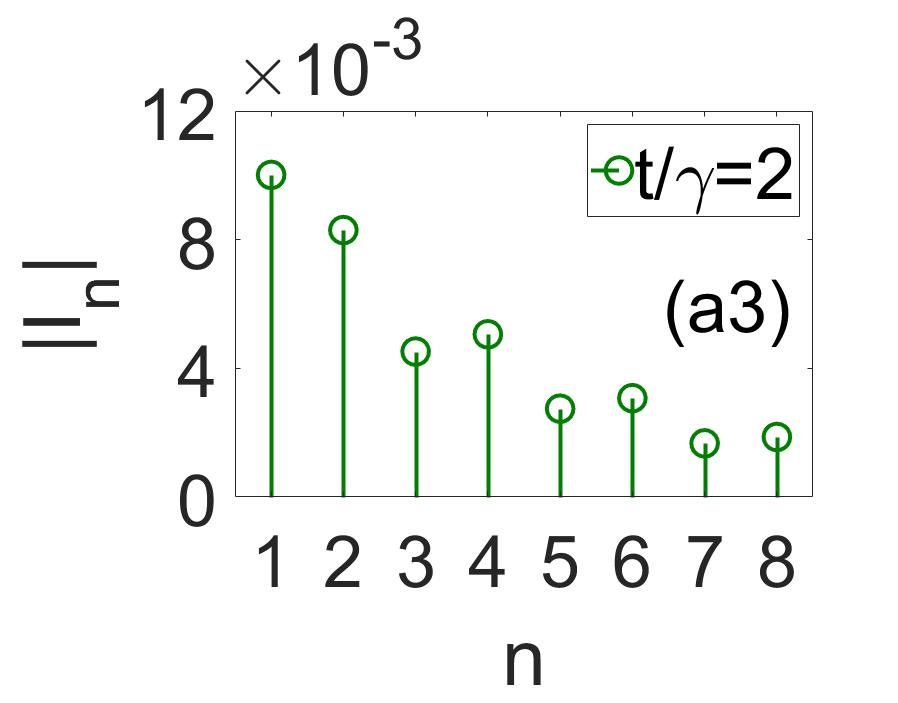}
    \includegraphics[width=4.27cm]{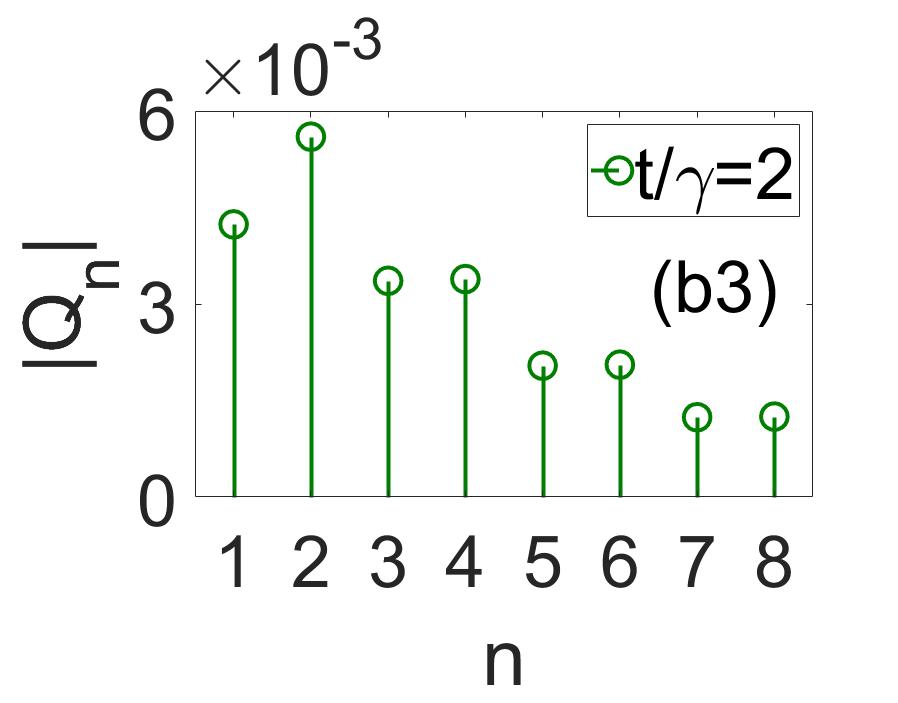}
    \caption{(i) Harmonics for particle current from the source: (a1) $t/\gamma=0.2$, (a2) $t/\gamma=1$, and (a3) $t/\gamma=2$; (ii) Harmonics for heat current from the source: (b1) $t/\gamma=0.2$, (b2) $t/\gamma=1$, and (b3) $t/\gamma=2$. Parameters used are $\gamma=0.05$, $\epsilon_d=8\gamma$, $\mu_D=-\mu_S=4\gamma$, $T_S=12\gamma$, $T_D=2\gamma$. Irrespective of $\phi$, the higher harmonics are non-zero for $t/\gamma=1$ and $t/\gamma\ge1$.  }
    \label{harmonics}
\end{figure}
The above observations can be further understood with the help of harmonic analysis of the particle and heat currents. One can analyze it as described below: The Aharonov-Bohm oscillations of charge current have a periodicity equal to $\Phi_0=h/e$. However, higher harmonics with periodicity $\Phi_0/n$ can also be observed with $n\ge 2$. These higher harmonics can be related to the circulation of electrons $n$ times in the triple-dot ring \cite{D'Anjou2013}.\\
 We define the harmonics for the particle current as
\begin{equation}
    I_n=\int_{0}^{2\pi}d\phi\hspace{0.1cm}I(\phi)e^{in\phi}.
\end{equation}
Similarly, the harmonics for the heat current can be defined as
\begin{equation}
    Q_n=\int_{0}^{2\pi}d\phi\hspace{0.1cm} Q(\phi)e^{in\phi},
\end{equation}
where $\phi=2\pi\Phi/\Phi_0$. Here $I(\phi)$ and $Q(\phi)$ are the expressions for particle current and heat current from the source as discussed in Eqs. (\ref{eq18}) and (\ref{eq19}), respectively.\\
\indent
   Here we observe that only the first two harmonic modes (mostly the second harmonic) are dominating for the strong coupling case ($t<\gamma$) as shown in Fig. \ref{harmonics}(a1) and \ref{harmonics}(b1). In this regime, the transmission has a single peak at $\omega=\epsilon_d$ i.e., only one state on a resonance that participates in the transport. This leads to the high efficiency of the system but low power output. Since the electrons coming from the source can circulate in the loop only once or twice and exit from the loop to the drain. We find that the $t\sim\gamma$ regime is the optimal regime for the operation of the thermoelectric heat engine. Since some of the higher harmonics ($n>2$) are also contributing to this regime. Here all three states contribute to the transport. The output power is enhanced as the magnitude of the charge and heat current harmonics increases as shown in Figs. \ref{harmonics}(a2) and \ref{harmonics}(b2), respectively. There is a balance between efficiency and output power in this region. In other words, the time scale of internal coherence dynamics of the triple dot should be comparable to the rate at which the triple dot state decays to the source and or drain terminal to achieve higher power output. In the weak coupling region, $t>\gamma$, all the higher harmonics contribute considerably. Hence the efficiency and output power are reduced as compared to the $t\sim\gamma$ regime. Although the output power is quite large as compared to the $t<\gamma$ regime. We can see that the magnitude of the higher harmonics is still comparable, but the first harmonics of the charge current are greater in magnitude than that of the heat current. This lag between harmonics results in lower efficiency for $t>\gamma$ regime .\\
\begin{figure}[t]
    \centering
    \includegraphics[width=6cm]{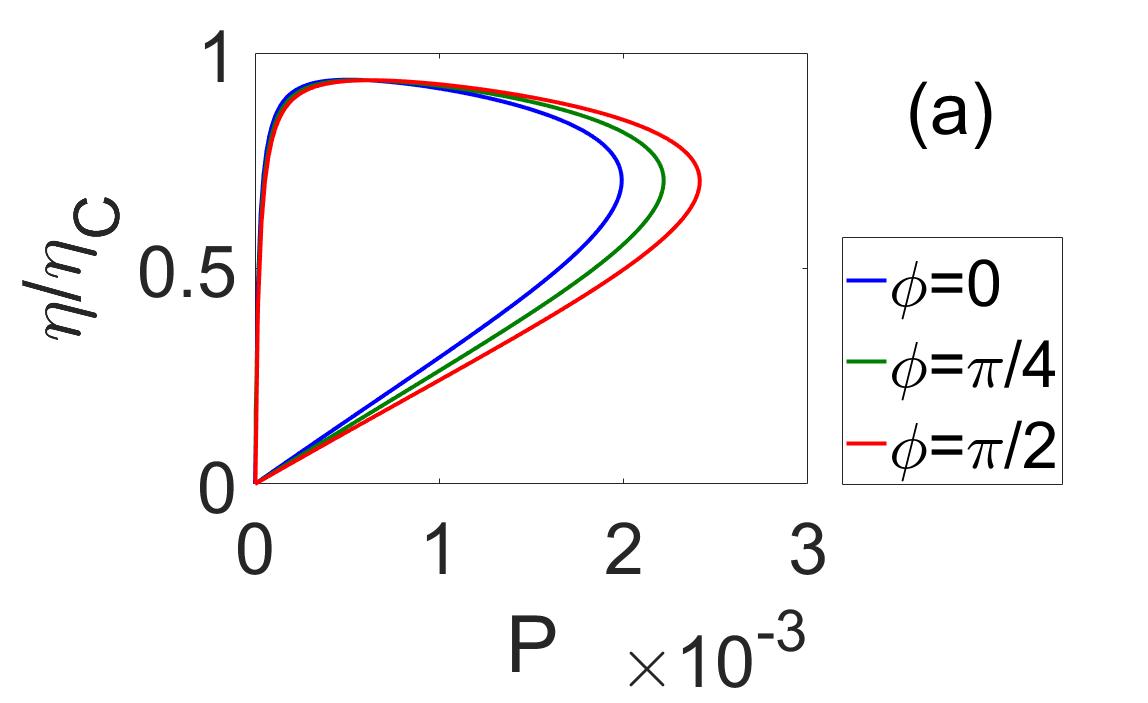}
    \includegraphics[width=6cm]{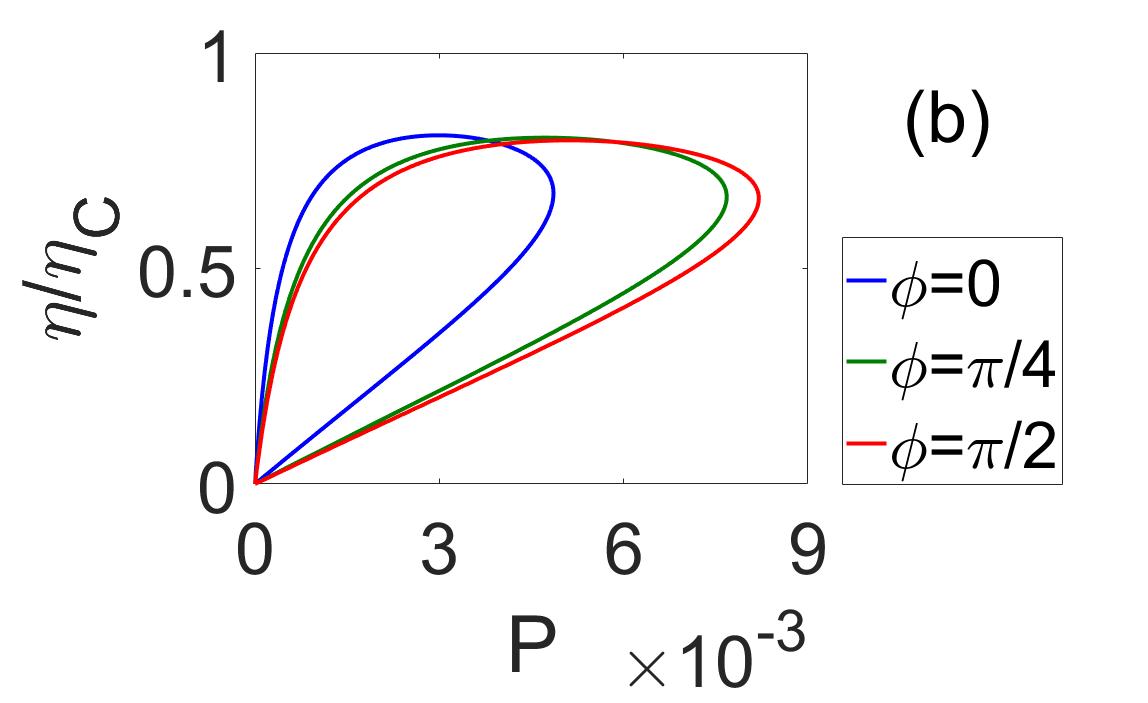}
    \includegraphics[width=6cm]{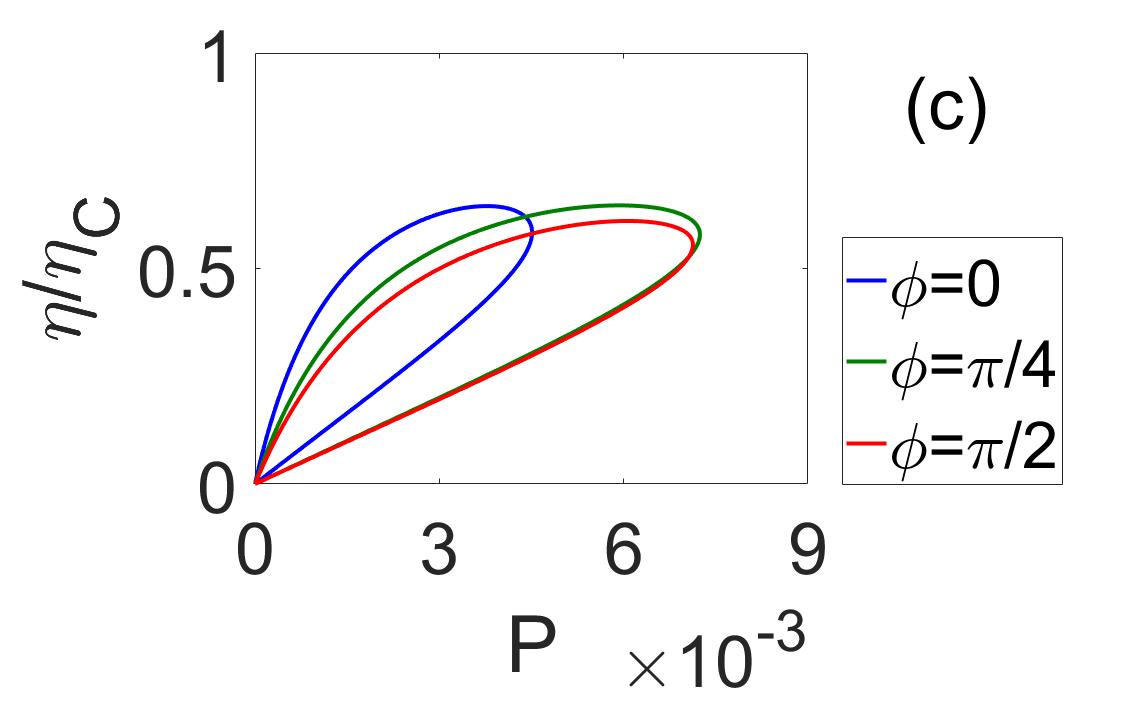}
    \caption{Power-efficiency diagram for different values of $\phi$ at (a) $t/\gamma=0.2$, (b) $t/\gamma=1$, (c) $t/\gamma=2$. Other parameters are $\gamma=0.05$, $\epsilon_d=8\gamma$, $\mu_D=-\mu_S$, $T_S=12\gamma$, $T_D=2\gamma$.}
    \label{power_effc_phi}
\end{figure}
Figure \ref{power_effc_phi} shows the trade-off between the output power and efficiency for different values of magnetic flux $\phi$ in three different regimes of $t$. It is observed that for all the three regimes of $t$, we get maximum output power at $\phi=\pi/2$ and the power is comparably less at $\phi=0$. This can be explained by the transmission function behavior that we observe as an antiresonance dip as shown in Figs. \ref{trans_less}(b1), \ref{trans_equal}(b1) and \ref{trans_large}(b1). Two symmetric resonance peaks around $\epsilon_d$ for $\phi=\pi/2$ open up more transport channels for charge and energy transport that give optimal power efficiency configuration.\\
\indent
From the above analysis, we can conclude that the presence of three resonance peaks with optimal separation between them is crucial for obtaining optimal output power and efficiency as in the case of $t\sim\gamma$. The presence of antiresonance dips in the $t>\gamma$ regime causes a reduction in both output power and efficiency. The harmonic analysis in Fig. \ref{harmonics} helps us understand the role of higher harmonic modes in enhancing thermoelectric performance. The contribution of higher harmonics is necessary for obtaining optimal output power but is not sufficient. When there is a more significant contribution from higher harmonic modes and the electrons spend more time circulating the loop rather than exiting the loop, then there is a reduction in efficiency and output power. We need a balance between the harmonics of both particle current and heat current to obtain optimal power output and efficiency. Also, we investigate the tunability of the thermoelectric performance of our model with the magnetic flux. The triple-dot exhibits maximum constructive interference at $\phi=\pi/2$ and we obtain maximum output power at $\phi=\pi/2$ for all the three regimes of $t$. With this analysis, we can conclude that $t\sim\gamma$ and $\phi=\pi/2$ are the optimal configurations to reach maximum output power and efficiency. In the following section, we study the power-efficiency behavior for a geometrically asymmetric setup with dissimilar dot-lead coupling strength.
\section{Beyond the wide-band limit approximation}\label{wbl}
\begin{figure}[t]
  \centering
  \includegraphics[width=6.2cm]{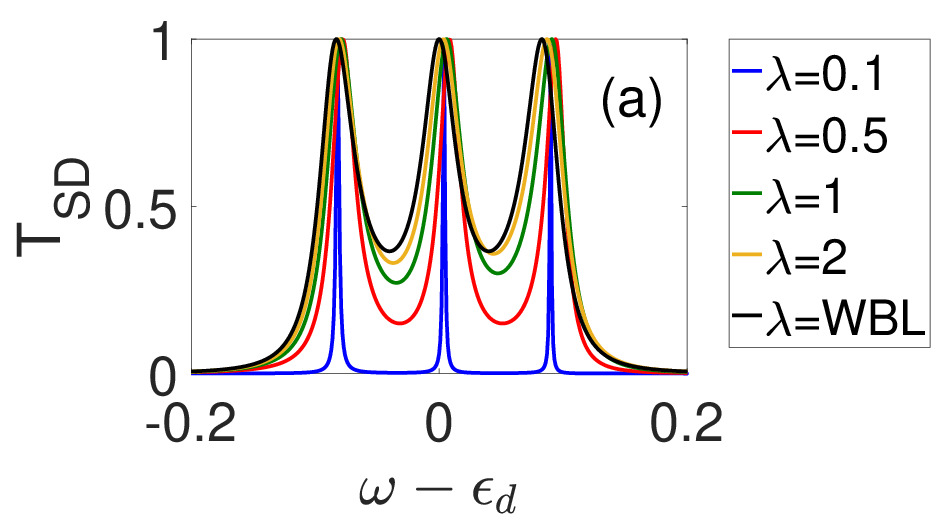}
  \includegraphics[width=6.2cm]{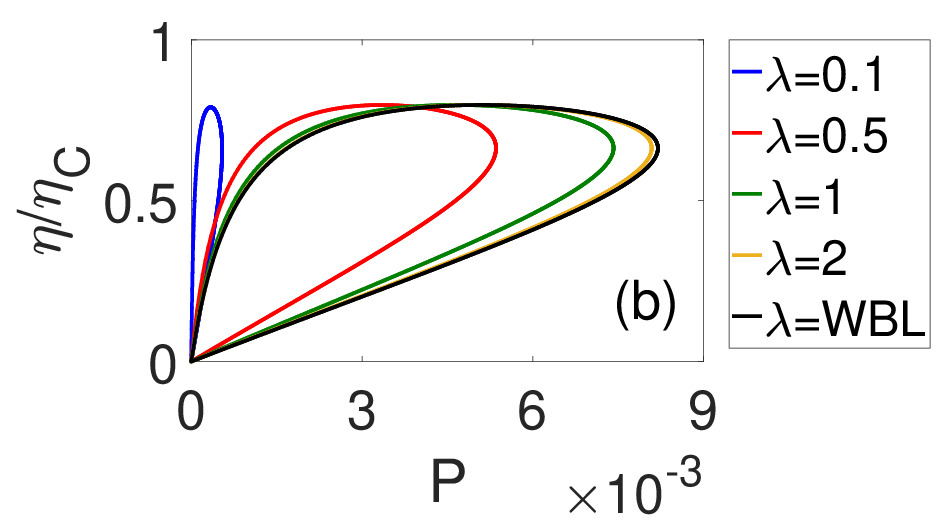}
  \caption{(a) Transmission function and (b) power-efficiency plot beyond the wide-band limit approximation. Parameters used are $t=\gamma=0.05$, $\epsilon_d=8\gamma$, $\mu_S=-\mu_D$, $T_S=12\gamma$, $T_D=2\gamma$, $\phi=\pi/2$. Here $\lambda =0.1=2\gamma$, $\lambda=0.5=10\gamma$,
$\lambda=1=20\gamma$, $\lambda=2=40\lambda$.}\label{wbl_fig}
\end{figure}
In the previous sections, we discuss the energy independent self-energies by taking the wide-band limit approximation (WBL). In this section we go beyond the WBL approximation and define the self-energies with minimal energy dependence as follows \cite{jctcWBL}:
\begin{equation}
    \Sigma^{\pm}_{\nu}(\omega)=\frac{1}{2}\bar{\Gamma}^{\nu}\frac{\lambda}{\omega\pm i\lambda}=\frac{1}{2}\bar{\Gamma}^{\nu}\Big(\frac{\lambda\omega}{\omega^2+\lambda^2}\mp i\frac{\lambda^2}{\omega^2+\lambda^2}\Big)
\end{equation}
where $\nu=S,D$ and $\lambda$ is a rescaling factor that allows us to approach the WBL as $\lambda\to\infty$. We define $\bar{\Gamma}^{\alpha}$ as
\begin{equation}
    \bar{\Gamma}^S=
    \begin{pmatrix}
        \gamma_s & 0 & 0\\
        0 & 0 & 0\\
        0 & 0 & 0
    \end{pmatrix},
    \,\,\,
    \bar{\Gamma}^D=
    \begin{pmatrix}
        0 & 0 & 0\\
        0 & 0 & 0\\
        0 & 0 & \gamma_d
    \end{pmatrix}
\end{equation}
The nonequilibrium Green's function is defined in Eq. (\ref{eqgreen}). Beyond the WBL approximation, the hybridization matrix becomes energy dependent and it is defined as
 \begin{equation}
     \Gamma^{\nu}(\omega)=i[\Sigma^{+}_{\nu}-\Sigma^{-}_{\nu}]=\Big(\frac{\lambda^2}{\omega^2+\lambda^2}\Big)\bar{\Gamma}^{\nu}.
 \end{equation}
 Using these above expressions we can define the transmission function from the source to the drain as
 \begin{equation}
     T_{SD}(\omega)=\text{Tr}[\Gamma^{S}G^+\Gamma^DG^-].
 \end{equation}
Figure \ref{wbl_fig} shows the plot for transmission function and power-efficiency behavior as we go beyond the WBL approximation towards the narrow band. As we move from the WBL approximation (i.e., $\lambda\to\infty$) towards the narrow band limit, the antiresonance dips in the transmission are more pronounced and we obtain three sharp delta-like resonance peaks for $\lambda=0.1$ as shown in Fig. \ref{wbl_fig}(a). These anti-resonance dips give rise to destructive interference channels causing the reduction in output power as evidenced in Fig. \ref{wbl_fig}(b). We obtain maximum output power in the WBL approximation and the output power reduces drastically for $\lambda=0.1$ although maximum efficiency is almost similar. These results support our claim that the three closely spaced sharp resonance peaks are necessary to obtain optimal power-efficiency. This is only observed in the wide-band limit. In the narrow-band limit, these peaks are far apart and the anti-resonance dips are pronounced leading to a reduction in power and efficiency.
\section{Disorder effects}\label{disorder}
\begin{figure}[t!]
  \centering
  \includegraphics[width=4.27cm]{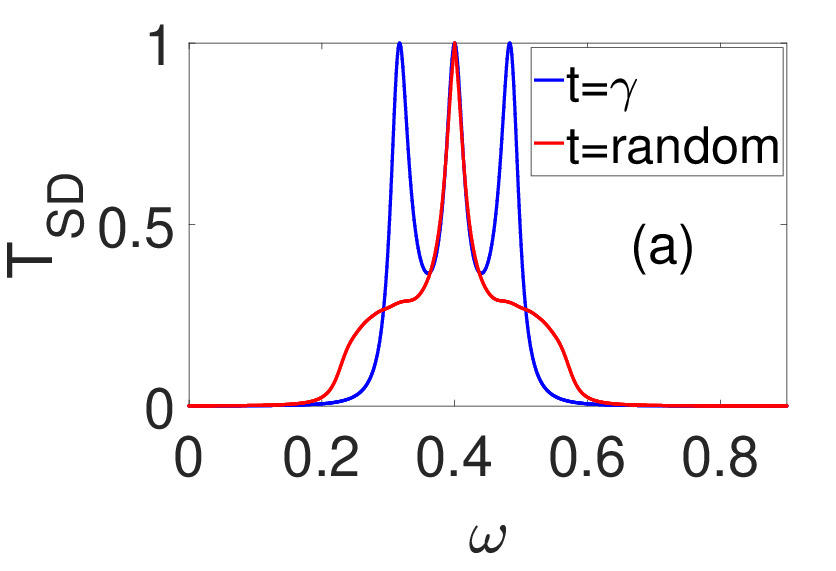}
  \includegraphics[width=4.27cm]{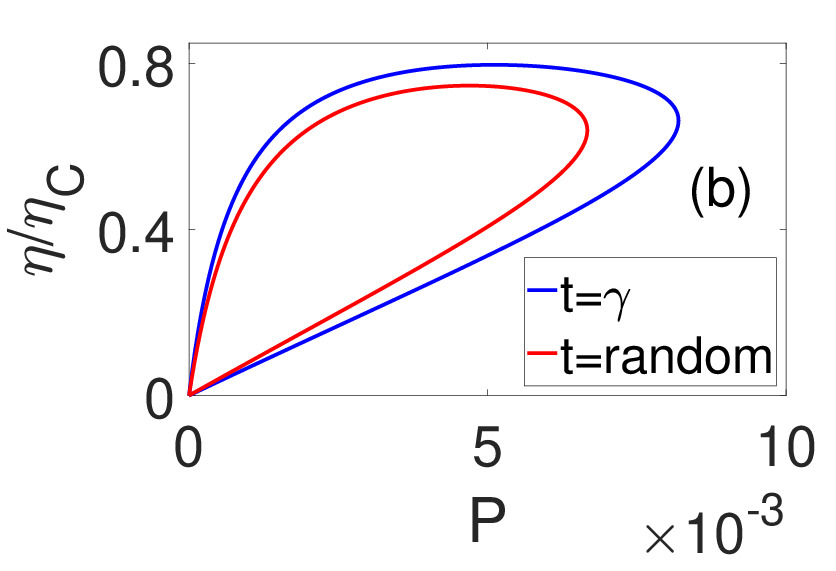}
  \includegraphics[width=4.27cm]{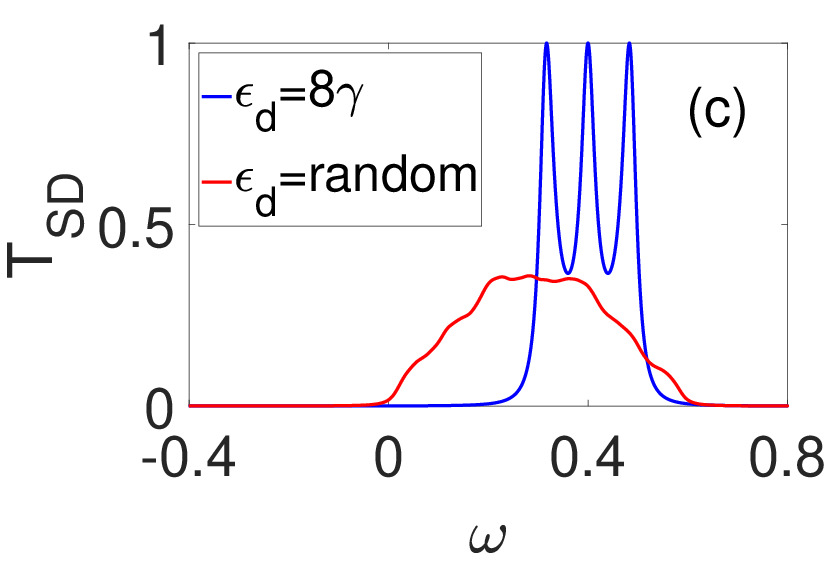}
  \includegraphics[width=4.27cm]{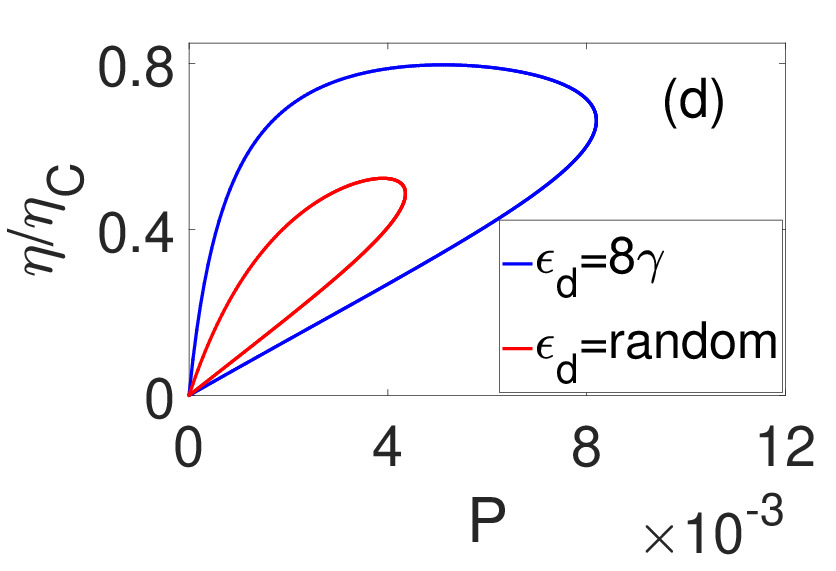}
  \caption{Plot of (a) transmission function as a function of energy and (b) power-efficiency plot with randomness in tunneling strength within the range of $t\in(0.02\gamma,2\gamma)$, keeping $\epsilon_d$ fixed at $8\gamma$. On the other hand we plot (c) transmission function versus $\omega$ and (d) power-efficiency plot with randomness in the on-site energies in the range of $\epsilon_d\in(2\gamma, 10\gamma)$, keeping $t=\gamma$. Other parameters used are $\gamma=0.05$, $\mu_S=-\mu_D$, $T_S=12\gamma$, $T_D=2\gamma$, $\phi=\pi/2$.}\label{random}
\end{figure}
The transport properties in nanoscale systems are often influenced by the disorder due to the impurity or the fluctuation of the parameters related to the nanostructures. Hence, the effect of disorder on transport properties and also on thermoelectric properties needs to be assessed for the thermoelectric applications of quantum interference-based nanoscale devices. One may observe two different kinds of disorder due to the impurity or fluctuation: (i) disorderliness in tunneling strength and (ii) disorderliness of on-site energies. First, we discuss the effect of a random change in the tunneling strength of each of the connecting legs in the central triple quantum dot region and calculate the transmission spectra and their thermoelectric properties. The on-site energies are set to a constant value $\epsilon_d=8\gamma$. We introduce disorders into our system by taking uniformly distributed random values of tunneling strength and then doing the disorder average to calculate the steady-state observables i.e., particle currents, heat currents, etc. For our numerical simulation, we chose random values of tunneling strength within the range $t\in(0.02\gamma,2\gamma)$ and investigated the disorder effects on the performance of the thermoelectric heat engine. In a similar fashion, we can also take into account the effect of disorderliness by introducing a random shift to each of the on-site energies in the central region by picking up random values of $\epsilon_d$ from a uniform distribution in the range $\epsilon_d\in(2\gamma, 10\gamma)$, considering $t=\gamma$. In both cases, we took 1000 samples of our model system and analyzed the output power, efficiency, and the corresponding transmission functions. Figure \ref{random} shows the effect of disorders on the transmission function and on the power-efficiency trade-off of our model AB heat engine. Disorder effects cause a reduction in both the output power and efficiency of the triple-dot AB heat engine, as expected in the previous studies \cite{disorder1,disorder2}. It is interesting to note that the central peak in transmission remains intact even in the presence of the disorder in tunneling strength $t$. The other two sharp resonance peaks are flattened (see Fig. \ref{random}(a)), yet the minimum transmission is $1/2$. Therefore, the power and efficiency are slightly reduced. In the case of the random disorder in on-site energies, the transmission function is flat and hence there is a substantial reduction in power-efficiency as shown in Fig. \ref{random}(c).
%It is observed that the disorderliness due to $\epsilon_d$ is much more detrimental compared to the randomness of tunneling strength. This can be explained by the characteristics of the corresponding transmission functions. For the randomness in $\epsilon_d$, the three resonance peaks of the transmission function disappear and it becomes a single broad distribution. On the other hand, the central resonance peak survives, but the other two peaks are smudged for the disorderliness of tunneling strength. As a result the power efficiency drastically reduces for random $\epsilon_d$ (approximately 50\%), but the randomness of $t$ lessens the power efficiency by an amount of 10\% compared to the pristine case.}
\section{Power-efficiency trade-off for asymmetric setups}\label{geometry}
Real experimental systems can exhibit asymmetries arising from imperfections in fabrications and/or disorder. Therefore, it is instructive to investigate the effects of geometric asymmetries on power-efficiency behavior.  We can introduce asymmetry to our system by introducing asymmetric dot-lead coupling strength ($\gamma_S\ne\gamma_D$). Let us define the asymmetric parameter as $x=\gamma_S/\gamma_D$. We discuss asymmetric behavior in the following section.
\subsection{Asymmetric dot-lead coupling strength}
\begin{figure}[t!]
    \centering
    \includegraphics[width=6.2cm]{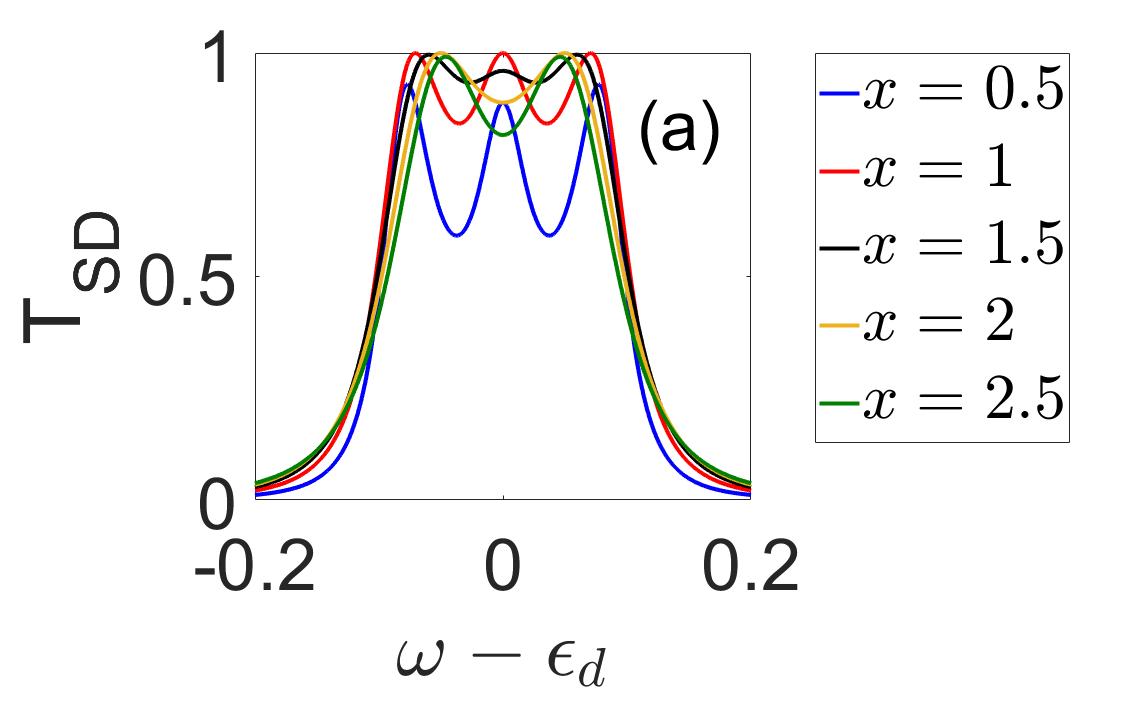}
    \includegraphics[width=6.2cm]{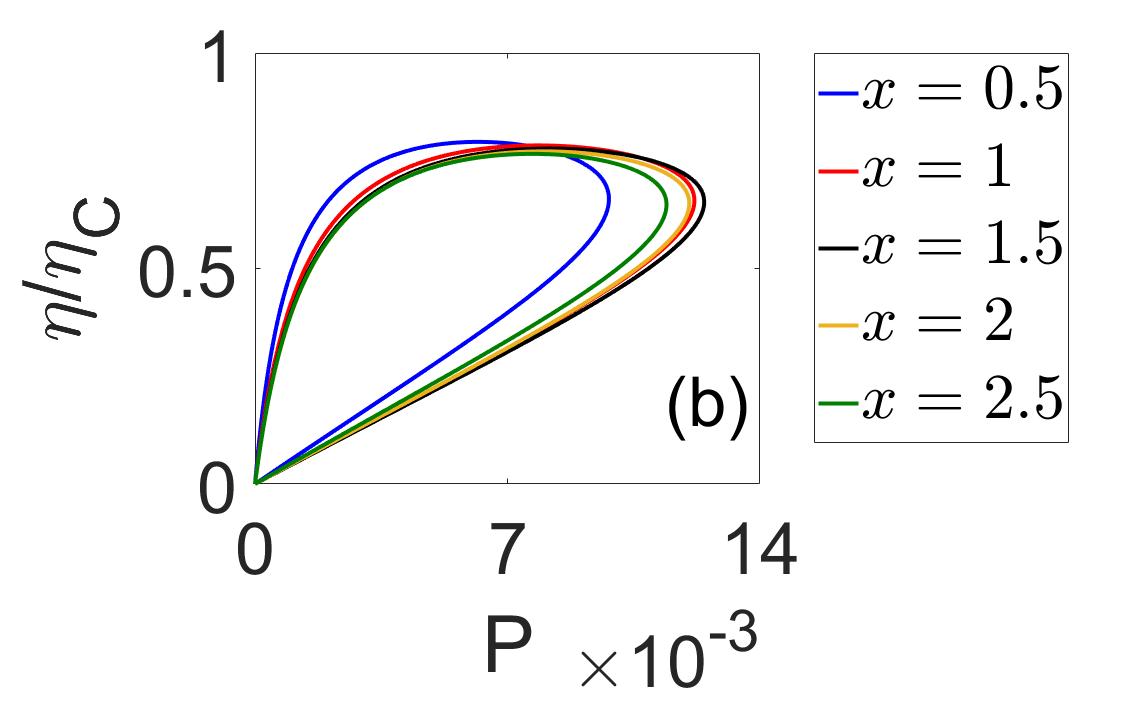}
    \includegraphics[width=6.2cm]{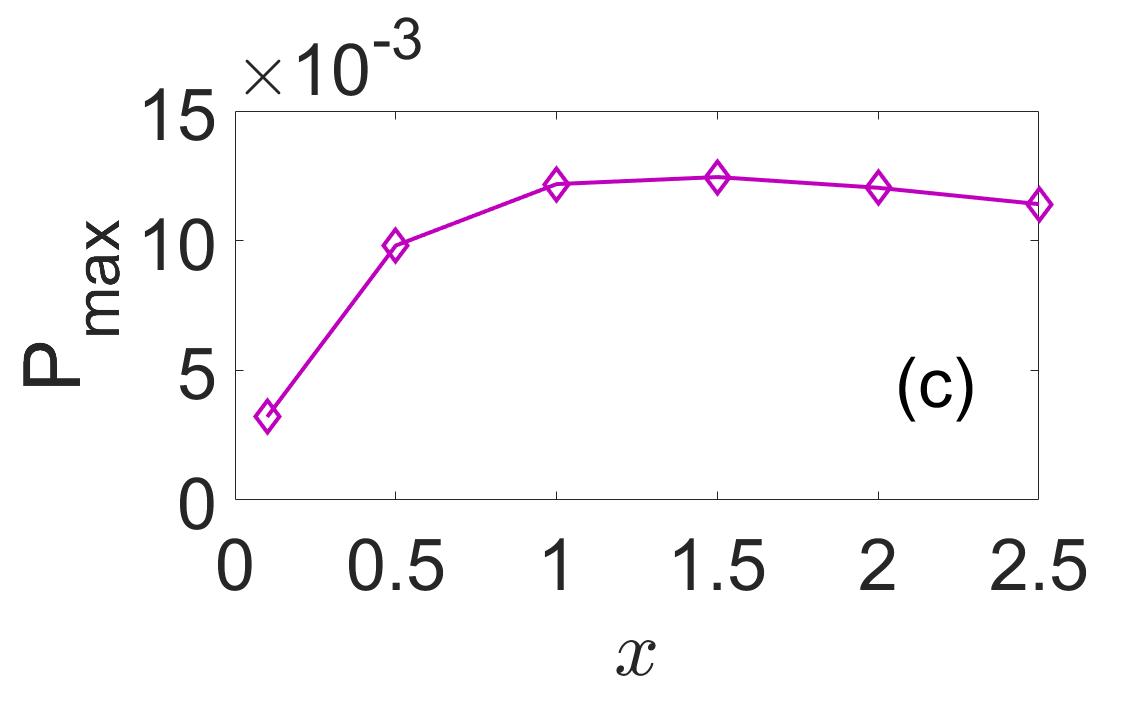}
    \caption{(a) Transmission function, (b) Power-efficiency diagram for asymmetric dot-lead coupling, and (c) maximum output power ($P_{max}$) as a function of asymmetric dot-lead coupling ratio. Parameters used $\gamma_D=0.1$, $t=0.5\gamma_D$, $\epsilon_d=4\gamma_D$, $\phi=\pi/2$, $\mu_D=-\mu_S$, $T_S=6\gamma_D$, $T_D=\gamma_D$. }
    \label{power_effc_asym}
\end{figure}
\begin{figure}[t!]
    \centering
    \includegraphics[width=4.27cm]{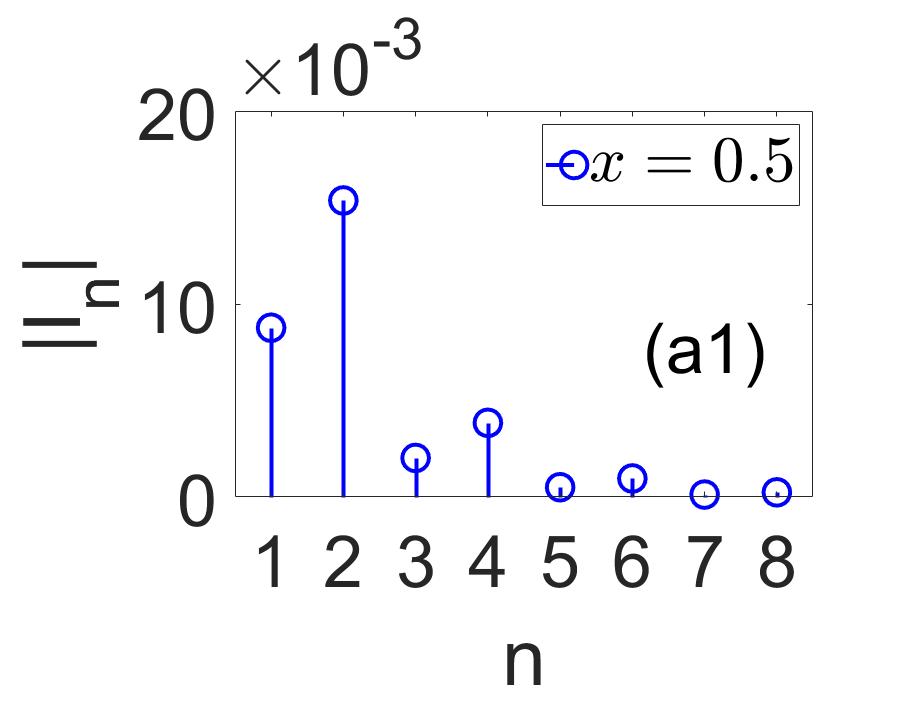}
    \includegraphics[width=4.27cm]{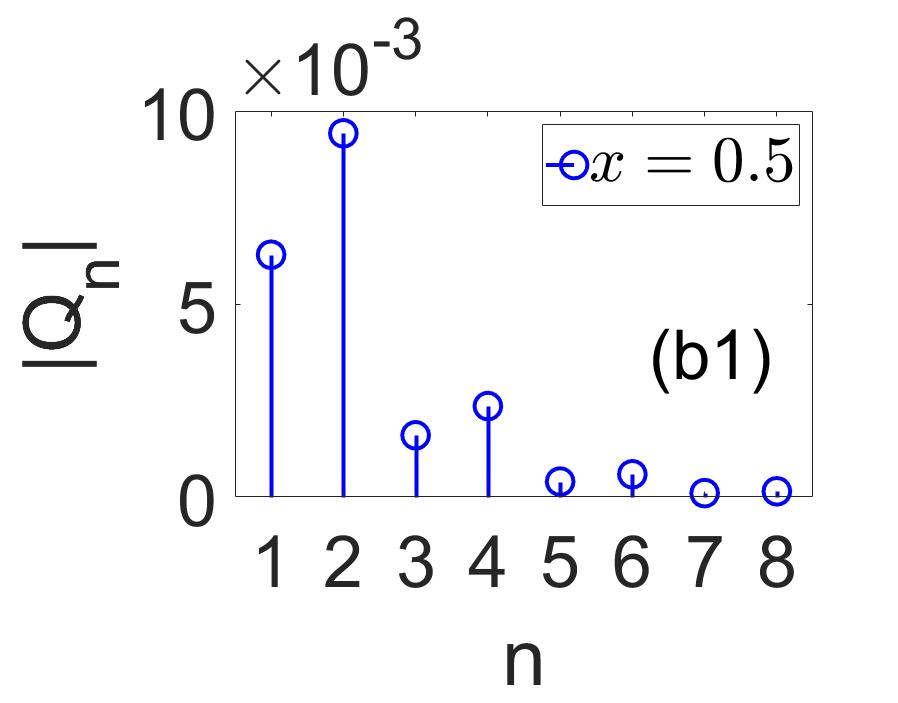}
    \includegraphics[width=4.27cm]{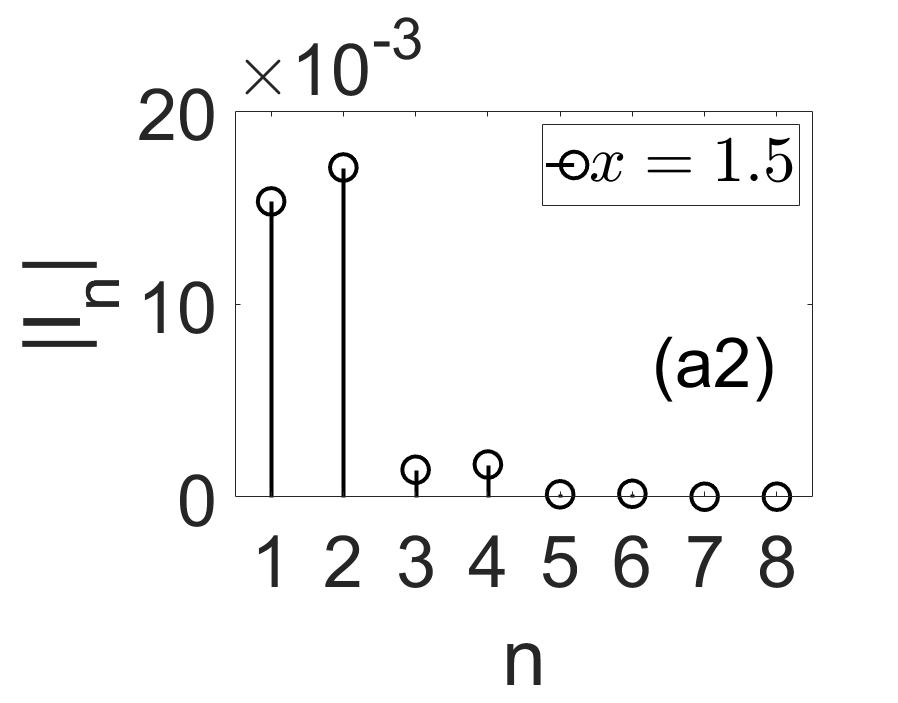}
    \includegraphics[width=4.27cm]{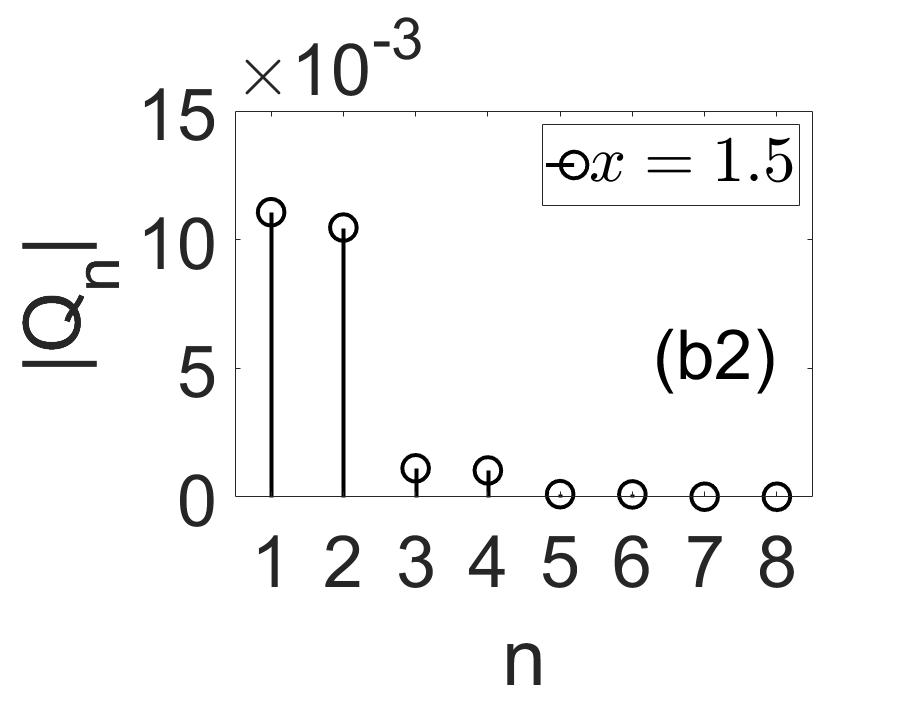}
    \includegraphics[width=4.27cm]{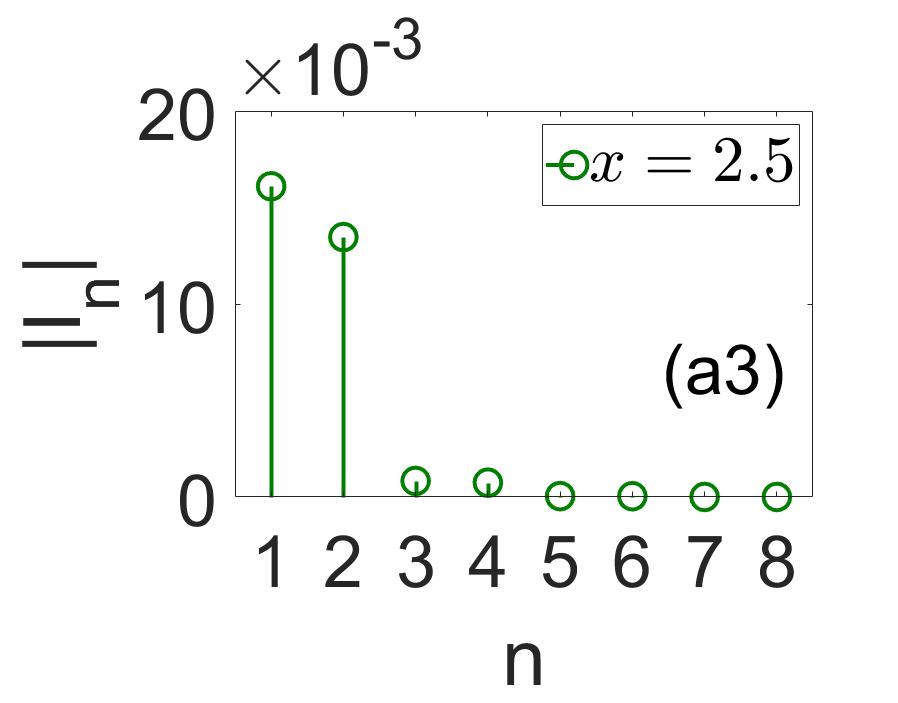}
    \includegraphics[width=4.27cm]{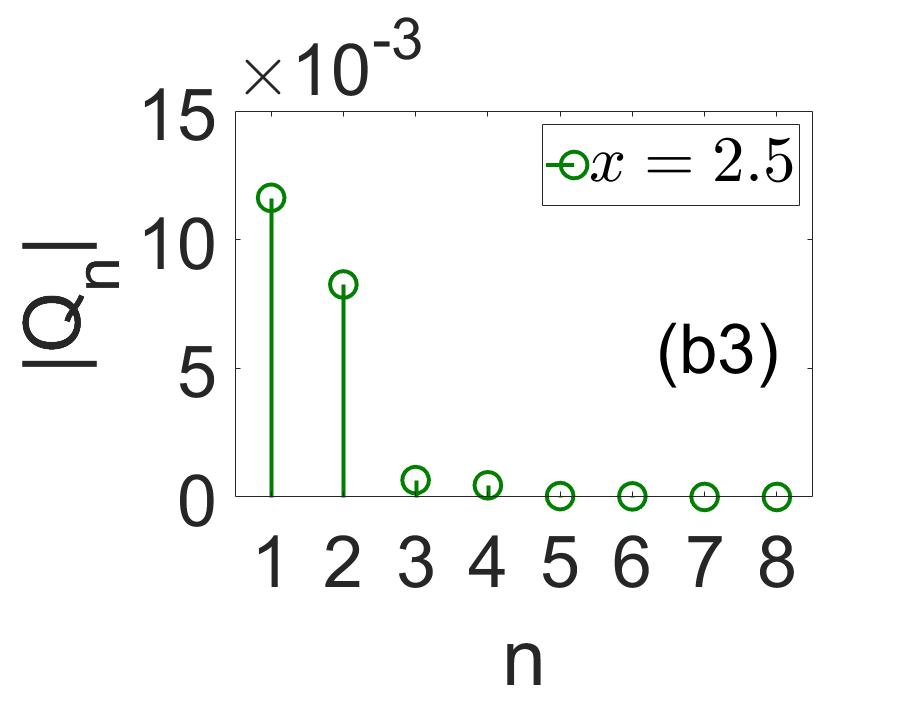}
    \caption{(i) Harmonics for particle current from the source: (a1) $x=0.5$, (a2) $x=1.5$, and (a3) $x=2.5$; (ii) Harmonics for heat current from the source: (b1) $x=0.5$, (b2) $x=1.5$, and (b3) $x=2.5$ for asymmetric dot-lead coupling. Parameters used $\gamma_D=0.1$, $t=0.5\gamma_D$, $\epsilon_d=4\gamma_D$, $\phi=\pi/2$, $\mu_D=-\mu_S=2\gamma_D$, $T_S=6\gamma_D$, $T_D=\gamma_D$.}
    \label{asym_harmonics}
\end{figure}
The power and efficiency of our heat engine model, also depend on the asymmetric dot-lead coupling strength (see Fig.\ref{power_effc_asym}).
For the AB  thermoelectric heat engine, it seems that an asymmetric coupling between the source-system, $\gamma_S$, and drain-system, $\gamma_D$, is also helpful to manipulate either efficiency or the power or both. Figure \ref{power_effc_asym} shows the transmission function and power-efficiency behavior for different asymmetric ratios $x$. Considering fixed values of the other parameters, we can achieve maximum efficiency up to 0.8 of Carnot efficiency by adjusting the asymmetric parameter $x$. By increasing the asymmetric parameter the maximum output power ($P_{max}$) increases from $x=0.5$ up to $x=1.5$ and then $P_{max}$ decreases by further increasing the asymmetric parameter for $x=2$ and $x=2.5$ as evidenced in Fig.\ref{power_effc_asym}(b) and Fig.\ref{power_effc_asym}(c). This can be explained by the transmission functions in Fig.\ref{power_effc_asym}(a). For $x=0.5, 1, 1.5$, there are three resonance transmission peaks. As the antiresonance dips are more significant for $x=0.5$, the maximum output power is less as compared to $x=1$ and the antiresonance dips are least significant for $x=1.5$ at which we obtain maximum output power $P_{max}=0.0125$. For $x=2$ and $x=2.5$, the three resonance transmission peaks merge to form two peaks causing a reduction of output power with antiresonance dips being more significant for $x=2.5$ which further reduces the maximum output power. So, by observing Fig. \ref{power_effc_asym}, we can conclude that whenever there are three peaks and the antiresonance dips are not much significant (i.e., $x=1.5$), we will get maximum output power. Figure \ref{asym_harmonics} shows the harmonic modes for particle current and heat current at different asymmetric parameters i.e., $x=0.5$, $x=1.5$, and $x=2.5$. This procedure does not reflect the absolute optimal values of power and efficiency as a function of all parameters of interest in possible experimental realizations, rather it demonstrates the high tunability of the AB interferometer as a heat engine to achieve a high amount of power and efficiency in the quantum regime.
\section{Conclusion and Discussion}\label{conclusion}
To conclude, we demonstrated the transmission engineering of a triple-dot Aharonov-Bohm interferometer to give an optimal power-efficiency configuration. We find that the higher-order interference patterns from multiple paths winding around the loop are necessary but insufficient for the optimal power-efficiency configuration. If there is a sufficiently strong hybridization of triple-dot states such that the time-scale of internal coherent dynamics (dictated by $1/t$) is comparable to the dot-lead hybridization strength, then we obtain three transmission peaks with transmission probability one and the two dips with the transmission probability one-half, then there is a maximally constructive interference effect that boosts the power-efficiency. In the present set-up, this condition is satisfied for $\phi=\pi/2$. Further, by tuning $\gamma=t$, the three transmission peaks can be brought closer in energy. The central peak occurs at $\omega=\epsilon_{d}$, and the other two peaks occur at $\omega=\epsilon_d\pm \frac{1}{2}\sqrt{12t^2-\gamma^2}$. For $\gamma=t$, $\omega=\epsilon_{d}\pm \frac{1}{2}\sqrt{11}t$ i.e peaks are closed for this case, and hence the constructive interference is achieved. By increasing $t$ further, we observed that the peaks move apart and the anti-resonance dips start to develop thereby reducing the power and efficiency. We also examined the effects of set-up asymmetry and observed that low to moderate asymmetry can enhance the power output.\\
\indent
Let us further proceed with our discussion by including some of the intriguing features of our phase-tunable AB quantum heat engine. The first question one may ask is how the electron-electron interaction in the nanostructure may affect the performance of the engine. In mesoscopic quantum-dot structures, the electron-electron interactions are not screened and therefore have to be taken into account beyond the mean-field description. This is a tedious task. But we can draw some qualitative insights into their effect on higher harmonics. First of all, in the non-interacting case, the charge and the heat currents are even functions of AB phase $\phi$ (owing to Onsager symmetry), and so $\cos(\phi)$ terms and their powers appear in the transmission probability. Incorporating the electron-electron interactions will break the even symmetry of charge and heat current. This will give rise to $\cos(\phi)$ and $\sin(\phi)$ terms which may suppress the harmonics beyond $n=2$. This has already been observed for a Coulomb blockade and infinite bias case \cite{Jin}. We expect that the suppression of higher harmonics may also suppress the power-efficiency. But irrespective of many-body interactions and their form, we would like to point out that an optimal structure of the transmission function is necessary so that maximal constructive interference can boost the power-efficiency of the engine.\\
\indent
 The response of the present thermoelectric heat engine turns out to be significant. Particularly, the quantum device based on an AB interferometer can produce a sizable thermopower. It is about one order of magnitude larger compared to that of the same kind of 3-dot interferometer operating in the linear regime \cite{Haack2019, Haack2021}. Moreover, thermoelectric efficiency at the maximum power of the heat engine is somewhat large and we obtain values as high as 80 percent of the Carnot efficiency in the present non-linear regime. Another aspect of the present study on this AB heat engine is that it can have great tunability either by changing magnetic flux, external gate voltage, or temperature bias. This will enable us to provide electrostatic-driven control of charge and heat current and its thermoelectric response under different relevant physical conditions achievable from the experimental point of view (as discussed below). \\
\indent
It is well known that the Onsager matrix in the linear response gives a complete
characterization of the transport properties. Such characterization
was investigated in Refs.\cite{brandner,Benenti1,Benenti2} and the enhancement in
efficiency was attributed to the breaking of time-reversal symmetry. Similar effects have been observed in other systems where the time-reversal symmetry is broken by an external drive \cite{zhou}. Henceforth, one natural question may arise: how the power-efficiency of our set-up be modified in the presence of elastic or inelastic scattering in the nonlinear regime?  In the nonlinear regime, Onsager symmetry is
broken in general (non-linear electrical and thermal conductance are not
even functions of magnetic field) \cite{ABring4,ABring5}. In our current set-up, we performed
preliminary simulations using the Buttiker voltage probe framework beyond
the linear response regime. We find that for low to moderate coupling strengths to the probe, the
transmission peaks are broadened thereby reducing output power and efficiency slightly. Thus, in the presence of low to moderate dephasing and inelastic scattering strengths, the presented thermoelectric effects are robust. Moreover, the parameters giving optimal performance are largely unaffected by the dephasing effect and inelastic scattering effect. Our observations are confirmed by the previous results \cite{quantumphase4,Sanchez1,Sanchez2,Sanchez3}. With the help of our previous study \cite{BANDYOPADHYAY2021114786}, we have seen that only the even harmonic modes are present for the charge current in the presence of an elastic scattering process. More investigations are underway and will be published elsewhere.\\
\indent
Although we have a toy model to understand the effect of quantum coherence on the optimal power-efficiency of AB heat engine, we may point out how this may be achieved in the laboratory. First, one may take into account that triangular triple quantum dots can be fabricated in various ways \cite{rogge,chen} to study different phenomena. Briefly one can mention that the dots, the barriers between the dots, and the coupling between the central structure to the source or drain can be controlled by electronic gates which modify the potential landscape of a 2-dimensional electron gas (2DEG). These gates can be generated by electron beam lithography \cite{noiri} or local anodic oxidation \cite{rogge} on top of an epitaxially grown 2DEG semiconductor heterostructure like GaAs/AlGaAs heterostructure. One may use top heaters for the effective thermal biasing of the QDs as mentioned in \cite{exptunits,glus} and then one may follow the measurement process as mentioned in \cite{exptunits}.\\
\indent
Our investigation establishes that one can trade-off the power-efficiency of a toy model of an AB thermoelectric heat engine by tuning the setup's quantum coherence and harmonic modes. The fascinating relationship between power-efficiency, external magnetic flux, gate voltage, temperature bias, and harmonic modes make this Aharonov–Bohm loop a prototypical platform for the execution of a distinctive class of phase-tunable thermoelectric quantum machines operating at cryogenic temperatures in the highly nonlinear regime. It is also observed that the symmetric triple-dot setup provides more tunability to control the thermoelectric response compared to its asymmetric counterpart.  Finally, in the context of quantum technologies, our studies on the coherent structure of AB interferometer provide a useful way to manipulate thermodynamic operational modes of quantum thermoelectric heat engines and such kind of setup might be at the core of several innovative thermoelectric quantum devices.\\
\section*{Acknowledgements}
We are grateful to Professor D. Segal for her valuable suggestions. J. B. acknowledges the financial support received from IIT Bhubaneswar in the form of an Institute Research Fellowship. M. B. gratefully acknowledges financial support from the Department of Science and Technology (DST), India under the Core grant (Project No. CRG/2020//001768). B. K. A. acknowledges the MATRICS grant (MTR/2020/000472) from SERB, Government of India and the Shastri Indo-Canadian Institute for providing financial support for this research work in the form of a Shastri Institutional Collaborative Research Grant (SICRG).
\appendix
\section{Equations of Motion}\label{appendixA}
The model setup of the triple-quantum-dot AB interferometer has been discussed in Section \ref{sec2}. We now solve this model and calculate the observables in the non-equilibrium steady-state. Since the model is noninteracting, we can use the Nonequilibrium Green's Function (NEGF) approach to calculate its steady-state characteristics \cite{meir1992, wangNEGF}. The NEGF technique has been widely used in the past years for investigating the transport properties in mesoscopic systems and molecular junctions \cite{textbook}. We follow the equation of motion approach for the derivations \cite{Dhar2006}. In this method, we solve the Heisenberg equations of motion (EOM) for the bath variables and then substitute them back in the EOM for the subsystem (dots) variables. We obtain a quantum Langevin equation (QLE) for the subsystem as follows:
\begin{equation}\label{eq6}
    \begin{split}
      \frac{d\hat{d}_i(t)}{dt}=-i\Big[\epsilon_i\hat{d}_i+\sum_{j\ne i} t_{ij}\hat{d}_j e^{i\phi_{ij}}\Big]
         -i\sum_{\nu=S,D}\hat{\eta}_i^{\nu}(t)\\
         -i\sum_{j,\nu=S,D}\int_{t_0}^{t}\Sigma_{i,j}^{\nu,+}(t-t')
         \hat{d}_j(t')dt'.
    \end{split}
\end{equation}
Here we use the indices $i=1,2,3$ to identify the three dots. The terms $\hat{\eta}_i^S$ and $\hat{\eta}_i^D$ are referred to as the noise induced on the subsystem by the source and drain, respectively and they are expressed as
\begin{equation}\label{eq7}
    \begin{aligned}
        \hat{\eta}_i^S=i\sum_k V_{i, k}^S\hspace{0.1cm}g_{Sk}^{+}(t-t_0)\hat{c}_{Sk}(t_0),\\
    \hat{\eta}_i^D=i\sum_k V_{i, k}^D\hspace{0.1cm}g_{Dk}^{+}(t-t_0)\hat{c}_{Dk}(t_0).
    \end{aligned}
\end{equation}
The retarded Green's functions of the isolated reservoirs are given by
\begin{equation}\label{eq8}
    \begin{aligned}
        g_{Sk}^{+}(t)=-ie^{-i\epsilon_{Sk}t}\theta(t),\\
        g_{Dk}^{+}(t)=-ie^{-i\epsilon_{Dk}t}\theta(t).
    \end{aligned}
\end{equation}
For the initial condition, we take factorized states for the total density matrix $\rho_T(t_0)=\rho_S\otimes\rho_D\otimes\rho(t_0)$, with empty dots and reservoirs prepared in a grand canonical state
\begin{equation}\label{eq9}
    \hat{\rho}_{\nu}=\frac{e^{-(\hat{H}_{\nu}-\mu_{\nu}\hat{N})/T_{\nu}}}{Tr[e^{-(\hat{H}_{\nu}-\mu_{\nu}\hat{N})/T_{\nu}}]},
\end{equation}
where $T_{\nu}$ and $\mu_{\nu}$ are the temperatures and chemical potentials of the Fermi sea with $\nu=S, D$. The state of the subsystem is denoted by the reduced density matrix $\rho$. Using the initial conditions we obtained the noise correlation as follows
\begin{equation}\label{eq10}
    \begin{aligned}
        \langle\hat{\eta}_i^{\dagger S}(t)
    \hat{\eta}_{i'}^{S}(\tau)\rangle=\sum_{k}V_{i,k}^{S^*}e^{i\omega_k(t-\tau)}V_{i',k}^S\hspace{0.1cm}f_S(\omega_k),\\
    \langle\hat{\eta}_i^{\dagger D}(t)
    \hat{\eta}_{i'}^{D}(\tau)\rangle=\sum_{k}V_{i,k}^{D^*}e^{i\omega_k(t-\tau)}V_{i',k}^D\hspace{0.1cm}f_D(\omega_k),
    \end{aligned}
\end{equation}
with the Fermi function $f_{\nu}=[e^{(\omega-\mu_{\nu})/T_{\nu}}+1]^{-1}$ for the reservoir $\nu=S,D$ with $\mu_{\nu}$ and $T_{\nu}$ be the corresponding chemical potential and temperature, respectively. In the Heisenberg picture, the expectation value of an observable $A$ can be obtained as
$\langle \hat{A}(t)\rangle=Tr_T[\rho_T(t_0)\hat{A}(t)]$, tracing over all degrees of freedom. The steady-state properties are obtained by taking the limits $t_0\rightarrow-\infty$ and $t\rightarrow\infty$. We can now take the Fourier transform of Eq. (\ref{eq6}) using the convolution theorem with the convention $\tilde{d}_i(\omega)=\int_{-\infty}^{\infty}dt\,d_i(t)e^{i\omega t}$
and $\tilde{\eta}_i^{\nu}(\omega)=\int_{-\infty}^{\infty}dt\,\eta_i^{\nu}(t)e^{i\omega t}$ and the result in matrix form is
\begin{equation}\label{eq11}
    \tilde{d}_i(\omega)=\sum_{j}G_{i,j}^{+}(\omega)\Big[\tilde{\eta}_j^{S}(\omega)+\tilde{\eta}_j^{D}(\omega)\Big].
\end{equation}
Here, the retarded Green's function is given by
\begin{equation}\label{eqgreen}
    G^{+}(\omega)=\Big[\omega I-\hat{H}_{TQD}-
    \Sigma_S^{+}(\omega)-\Sigma_D^{+}(\omega)\Big]^{-1},
\end{equation}
where $I$ is a $(3\times3)$ identity matrix and the advanced Green's function is given by the transpose conjugate of the retarded Green's function, $G^-(\omega)=[G^+(\omega)]^{\dagger}$. The self-energies are defined as:
\begin{equation}\label{eq13}
    \begin{aligned}
        \Sigma_S^{\pm}(\omega)=\sum_k V_{1,k}^S\,g_{S}^{\pm}(\omega)\,V_{1,k}^{S^*},\\
        \Sigma_D^{\pm}(\omega)=\sum_k V_{3,k}^D\,g_{D}^{\pm}(\omega)\,V_{1,k}^{D^*}.
    \end{aligned}
\end{equation}
Here, $g_{S}^{\pm}(\omega)$ and $g_{D}^{\pm}(\omega)$ are given by the Fourier transform of Eq. (\ref{eq8}). In the wide-band limit and when the density of states of the metallic lead is energy independent, the real part of the self-energy term vanishes. Then we can define the hybridization matrix from the relation $\Sigma^{+}=-i\Gamma/2$:
\begin{equation}\label{eqgamma}
    \Gamma_{i,i'}^{\nu}=2\pi\sum_{k,\nu}V_{i',k}^{\nu^*}V_{i,k}^{\nu}\, \delta(\omega-\omega_k).
\end{equation}
We may take $V_{i,k}^{\nu}$ as real constants, independent of the level index and reservoir state, resulting in $\Gamma_{1, 1}^S=\gamma_S$ and $\Gamma_{3, 3 }^D=\gamma_D$, where $\gamma_{\nu}$ (energy independent) describes the coupling between the dots and metallic leads. We consider degenerate dot energies $\epsilon_1=\epsilon_2=\epsilon_3=\epsilon_d$ and set $t_{ij}=t$ to obtain the retarded Green's function.

\section{Physical units}\label{appendixB}
Our paper uses the natural unit convention $\hbar=c=e=k_B$ for simulation. At this point, it is useful to express our results in terms of physical units. Thermoelectric heat
engine based on a quantum dot (QD) embedded into a semiconductor nanowire has been studied experimentally \cite{exptunits}. One can obtain physical units using the relation $\hbar\gamma=k_BT_a$, where $\gamma$ is the dot-lead coupling and $T_a$ is the average temperature. Considering $T_a=1\,\mathrm{K}$, we obtain $\gamma=1.3\times 10^{11}\,\mathrm{Hz}$ and based on this value of $\gamma$ we convert all other parameters to physical units. In simulations, we use the dot-lead coupling strength, $\gamma=0.05$ which translates into $\gamma=6.5\,\mathrm{GHz}$. The parameters $t=\gamma$, $\epsilon_d=8\gamma$, $T_S=12\gamma$, $T_D=2\gamma$, $\mu_D=-\mu_S=4\gamma$ translate to $t=4.3\,\mathrm{\upmu eV}$, $\epsilon_d= 34.4\,\mathrm{\upmu eV}$, $T_S= 0.6\,\mathrm{K}$, $T_D= 0.1\,\mathrm{K}$, and $\mu_D=-\mu_S= 17.2\,\mathrm{\upmu eV}$, respectively. The total magnetic flux enclosed by the triangular AB ring can be calculated from Eq. (\ref{eq5}) as $\Phi=\frac{\phi\Phi_0}{2\pi}$, where $\Phi_0=h/e$ is the flux quantum. The magnetic field is defined as, $B=\frac{\phi\Phi_0}{2\pi A}$, where $A$ is the area of the AB ring. For an area of an $80\,\mathrm{nm}$ equilateral triangle, the AB oscillation period is of $\frac{\Phi_0}{A}\sim 1.5\,\mathrm{T}$. So, for an AB phase of $\phi=\pi/2$ and $\phi=\pi/4$ the value of the magnetic field used is $B\sim 0.375\,\mathrm{T}$ and $\sim 0.1875\,\mathrm{T}$, respectively. For details on the values of the magnetic field to be used to observe AB oscillation see Ref. \cite{hod2008}.\\
\indent
In physical units, the Landauer- Buttiker formula for electric current from the source (S) is defined by
\begin{equation}
    I_{S}=\frac{e}{h}\int_{-\infty}^{\infty}d\omega\big[
    T_{SD}(\omega,\phi)f_{S}(\omega)-
    T_{DS}(\omega,\phi)f_{D}(\omega)\big].
\end{equation}
and the heat current from the source (S) is defined by
\begin{equation}
\begin{split}
    Q_{S}=\frac{1}{h}\int_{-\infty}^{\infty}d\omega(\omega-\mu_{S})
    \big[T_{SD}(\omega,\phi)f_{S}(\omega)\\
    -T_{DS}(\omega,\phi)f_{D}(\omega)\big],
\end{split}
\end{equation}
where $f_{\nu}$, ($\nu=S,D$), is the Fermi distribution function, $e$ is the charge of the electron and $h$ is Planck's constant. Now, the output power generated is given by $P=(\Delta\mu/e)I_S$, where $\Delta\mu=\mu_D-\mu_S$. In physical units, $\omega=\mathcal{O}(\gamma= 1.3\times 10^{11}\,\mathrm{Hz})$. To express the output power in physical units, the factor $\gamma=1.3\times 10^{11}\,\mathrm{Hz}=13.76\times 10^{-24}\,\mathrm{J}$ should be multiplied with both $\Delta\mu$ and $I_S$.
With this, we focus on Fig. \ref{power_effc_t} and express the output power in physical units for three different regimes i.e., (a) $t/\gamma<1$, (b) $t/\gamma=1$ and (c) $t/\gamma>1$. Considering Fig. \ref{power_effc_t}, for $t/\gamma<1$ regime, we obtain the maximum output power, $P_{max}=0.0024$ corresponding to an electric current from the source, $I_S=0.0073$ which translates to $P_{max}\sim 0.7\,\mathrm{fW}$ and $I_S= 0.024\,\mathrm{nA}$, respectively. Similarly, for $t/\gamma=1$ regime, the maximum output power, $P_{max}=0.0082$ corresponding to an electric current of $I_S=0.026$ translates to $P_{max}\sim 2.35\,\mathrm{fW}$ and $I_S= 0.086\,\mathrm{nA}$, respectively. For $t/\gamma>1$ regime, maximum output power $P_{max}=0.0071$ corresponds to electric current $I_S=0.0262$ translates to $P_{max}\sim 2.04\,\mathrm{fW}$ and $I_S= 0.087\,\mathrm{nA}$, respectively.\\
\indent
Now we consider Fig. \ref{power_effc_asym}(c) and obtain the maximum output power for different asymmetric dot-lead coupling ratios in physical units. We define the asymmetric parameter as $x=\gamma_S/\gamma_D$. For $x=0.5$, we obtain a maximum output power $P_{max}=0.0098= 2.81\,\mathrm{fW}$. Similarly, for $x=1$, $P_{max}=0.0122= 3.5\,\mathrm{fW}$;
for $x=1.5$, $P_{max}=0.0125= 3.58\,\mathrm{fW}$;
for $x=2$, $P_{max}=0.0120= 3.44\,\mathrm{fW}$;
for $x=2.5$, $P_{max}=0.0114=3.27\,\mathrm{fW}$.

\section{Thermovoltage and Seebeck coefficient}\label{seebeck}
\begin{figure}[b]
  \centering
  \includegraphics[width=6.2cm]{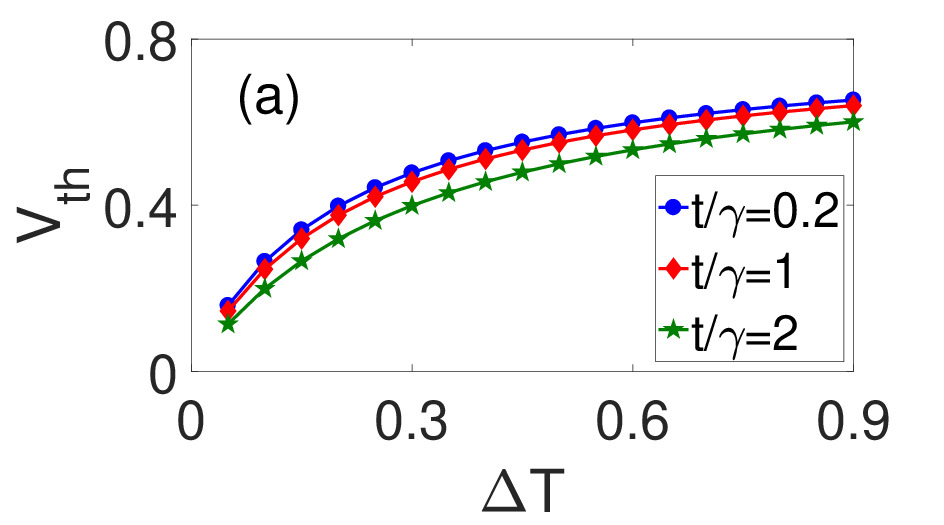}
  \includegraphics[width=6.2cm]{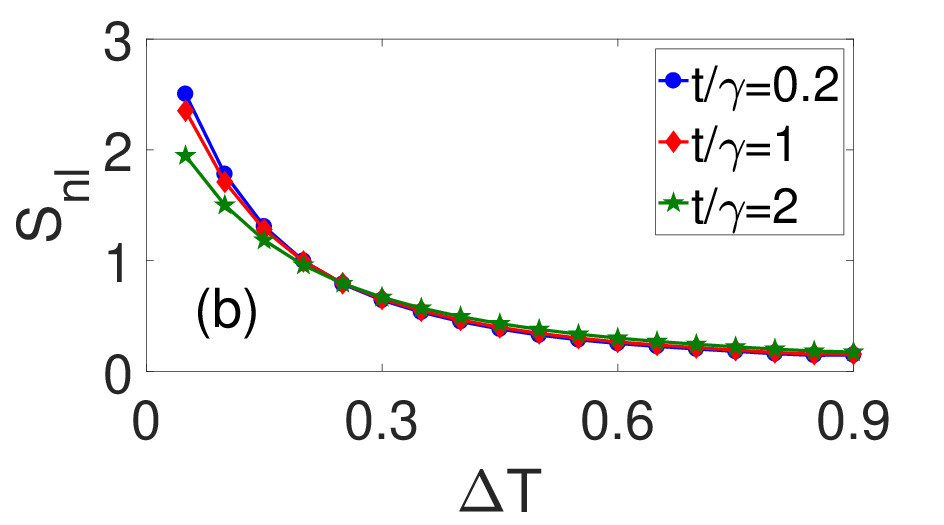}
  \caption{(a) Thermovoltage and (b) Seebeck coefficient as a function of $\Delta T$. Parameters used are $\gamma=0.05$, $\epsilon_d=8\gamma$, $\mu_S=-\mu_D$, $T_S=12\gamma$, $T_D=2\gamma$, $\phi=\pi/2$.}\label{thermo}
\end{figure}
Seebeck coefficient is a key parameter to measure the thermoelectric performance of a material and it tells us about the amount of voltage developed across a system in response to a given temperature bias at zero charge current i.e., $I_S=0$. Since the induced voltage is originated from the temperature bias we call it `thermovoltage ($V_{th}$)'. In the linear response regime, the Seebeck coefficient $S_l$ is defined as the ratio of the thermovoltage to the small temperature difference $\Delta T$ at zero charge current,
\begin{equation}\label{seebeck_linear}
  S_l=\frac{V_{th}}{\Delta T}\bigg|_{I_S=0}.
\end{equation}\\
In the nonlinear regime, it is the differential Seebeck coefficient $S_{nl}$ which is defined as the partial derivative of the thermovoltage with respect to $\Delta T$ \cite{Haack2021},
\begin{equation}\label{seebeck_nonlinear}
  S_{nl}=\frac{\partial V_{th}}{\partial\Delta T}\bigg|_{I_S=0}.
\end{equation}

Figure \ref{thermo} shows the thermovoltage and Seebeck coefficient developed across the system at different temperature bias $\Delta T$ for three different regimes of $t$. The thermovoltage increases with an increase in $\Delta T$ and is maximum for $t<\gamma$ regime and minimum in the $t>\gamma$ regime. However, the Seebeck coefficient is decreasing with $\Delta T$ and it is maximum for small $\Delta T$. The Seebeck coefficient $S_{nl}$ is maximum in The $t<\gamma$ regime and minimum in the $t>\gamma$ regime for small temperature biases. However, as we increase the temperature difference the Seebeck coefficient decreases rapidly and saturates towards high temperature biases for all the three regimes of $t$.

\end{document}